\documentclass[10pt,journal,compsoc]{IEEEtran}

\ifCLASSOPTIONcompsoc
  \usepackage[nocompress]{cite}
\else
  \usepackage{cite}
\fi
\ifCLASSINFOpdf
  \usepackage[pdftex]{graphicx}
  \DeclareGraphicsExtensions{.pdf,.jpeg,.png}
\else
\fi

\usepackage{amsmath}
\usepackage{hyperref}
\usepackage{enumitem}
\usepackage{algorithm}
\usepackage{xcolor}
\usepackage{color}
\usepackage{cite}
\usepackage{epstopdf}
\usepackage{booktabs}
\usepackage{multirow}
\usepackage{textcomp}
\usepackage{caption}

\hypersetup{hidelinks,
backref=true,
pagebackref=true,
hyperindex=true,
breaklinks=true,
colorlinks=true,
urlcolor=blue,
bookmarks=true,
bookmarksopen=false,
pdftitle={Title},
pdfauthor={Author}}

\def\eg{\textit{e.g.}}
\def\ie{\textit{i.e.}}

\def\etc{\textit{etc.}}
\newcommand*\rot{\rotatebox{90}}
\begin{document}
%
\title{Gender and Emotion Recognition from Implicit User Behavior Signals}
%
%
%
%

\author{Maneesh Bilalpur, 
        Seyed Mostafa Kia, ~\IEEEmembership{Member,~IEEE},
		Mohan Kankanhalli, ~\IEEEmembership{Fellow,~IEEE},
        Ramanathan Subramanian, ~\IEEEmembership{Senior Member,~IEEE}
\IEEEcompsocitemizethanks{\IEEEcompsocthanksitem Maneesh Bilalpur is with the School of Computing and Information, Univ. Pittsburgh, USA. (E-mail: mab623@pitt.edu) \protect
 \IEEEcompsocthanksitem Seyed Mostafa Kia is with the Donders Institute at Radboud University, Nijmegen, The Netherlands. (E-mail: s.kia@donders.ru.nl)\protect
\IEEEcompsocthanksitem Mohan Kankanhalli is with School of Computing at National University of Singapore, Singapore.~(E-mail: mohan@comp.nus.edu.sg)\protect
\IEEEcompsocthanksitem Ramanathan Subramanian is with the Indian Institute of Technology, Ropar. (E-mail: s.ramanathan@iitrpr.ac.in)\protect

}
}

\IEEEtitleabstractindextext{%
\begin{abstract}
This work explores the utility of \textbf{\textit{implicit}} behavioral cues, namely, \emph{Electroencephalogram} (EEG) signals and \emph{eye movements} for gender recognition (GR) and emotion recognition (ER) from psychophysical behavior. Specifically, the examined cues are acquired via low-cost, off-the-shelf sensors. 28 users (14 male) recognized emotions from unoccluded (\textit{no mask}) and partially occluded (\textit{eye} or \textit{mouth masked}) emotive faces; their EEG responses contained gender-specific differences, while their eye movements were characteristic of the perceived facial emotions. Experimental results reveal that (a) reliable GR and ER is achievable with EEG and eye features, (b) differential cognitive processing of negative emotions is observed for females and (c) eye gaze-based gender differences manifest under partial face occlusion, as typified by the \textit{eye} and \textit{mouth mask} conditions.  


\end{abstract}

}

\maketitle

\IEEEdisplaynontitleabstractindextext

%
\IEEEpeerreviewmaketitle

\IEEEraisesectionheading{\section{Introduction}\label{sec:introduction}}

\IEEEPARstart{G}ender human-computer interaction (HCI)~\cite{Susan2006} and Affective HCI~\cite{Picard1997} have evolved as critical HCI sub-fields, as it is critical for computers to \textit{appreciate} and \textit{adapt to} the user's gender and emotional state. Inferring users' \emph{soft biometrics} such as gender and emotion would benefit interactive and gaming systems in terms of a) visual and interface design~\cite{PassigL01,Czerwinski2002}, (b) game and product recommendation (via ads)~\cite{Zhang16,Homer2012}, and (c) provision of appropriate motivation and feedback for optimizing user experience~\cite{Schwark2013}. Gender recognition (GR) and emotion recognition (ER) systems primarily work with facial~\cite{Ng2012,joho2011looking} or speech~\cite{Li2013,lee2005toward} cues which are \textit{biometrics} encoding a person's identity. Also, they can be recorded without the user's knowledge, posing grave privacy concerns~\cite{privacy}. 

This work examines GR and ER from \textit{\textbf{implicit user behavioral signals}}, in the form of \textit{EEG brain signals} and \textit{eye movements}. Implicit behavioral signals are inconspicuous to the outside world, and cannot be recorded without express user cooperation making them privacy compliant~\cite{Campisi2014}. Also, behavioral signals such as EEG and eye movements are primarily \textit{anonymous} as little is known regarding their uniqueness to a person's identity~\cite{Yang2012}.  

\begin{figure}[t]
\centering
\captionsetup{justification=centering}
\includegraphics[scale=0.3]{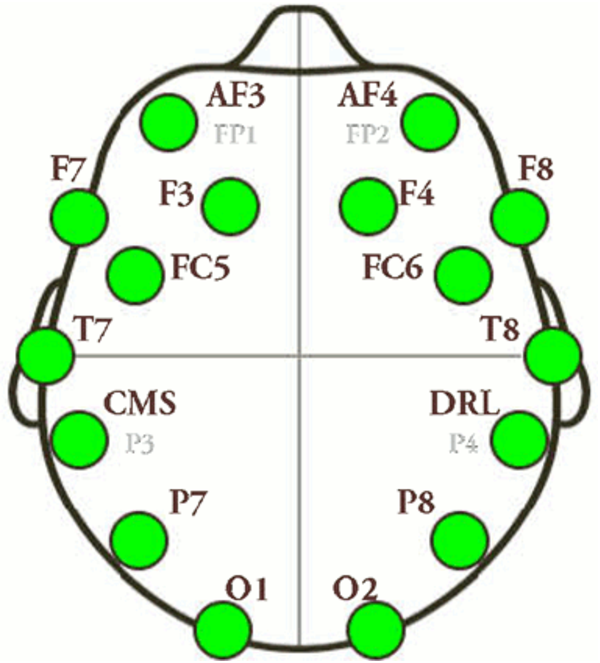}
\caption{\textbf{\textit{Emotiv Epoc+}} electrode configuration: The headset comprises 14 sensing plus two reference electrodes.}\label{Emotiv}
\vspace{-.4cm}
\end{figure}

Specifically, we attempt GR and ER using signals captured by commercial, off-the-shelf devices which are minimally intrusive, affordable, and  popularly used in gaming as input or feedback modalities~\cite{Vliet12,LimSW15}. The \textit{Emotiv} EEG wireless headset consists of 14 dry (plus two reference) electrodes having a configuration as shown in Fig.~\ref{Emotiv}. While being lightweight, wearable and easy-to-use, neuro-analysis with \emph{Emotiv} can be challenging due to relatively poor signal quality. Likewise, \textit{EyeTribe} is a low-cost eye-tracker whose suitability for research has been endorsed~\cite{dalmaijer2014low}. We show how relevant gender and emotion-specific information is captured by these low-cost devices via examination of event-related potential (ERP) and eye fixation patterns, and also through recognition experiments.

We set out to discover gender differences in human visual perception by designing a facial emotion recognition (FER) experiment. Males and females respond differently to affective information~\cite{Schwark2013,montagne,halljudith,maneesh2017acii}, and user eye movements are also characteristic of the perceived facial emotion~\cite{Tavakoli15,vassallo2009visual,wells2016identification,Subramanian2014,schurgin}, enabling stimulus ER. Our study performed with 28 viewers (14 males) confirms that women achieve superior FER, mirroring prior findings. Hypothesizing that enhanced female emotional sensitivity should reflect via their \emph{implicit} behavior, we examined EEG and eye-gaze patterns to find that (1) Stronger ERPs are observed for females while processing negative facial emotions, (2) Female eye-gaze is more focused on the eyes for the purpose of FER, and (3) Emotion and information-specific gender differences manifest starkly, enabling better GR under particular  stimulus conditions. 

Building on preliminary results~\cite{maneesh2017acii,Bilalpur2017}, our work makes the following research contributions: (a) While prior works have identified gender differences in emotional behavior, this is one of the first works to expressly perform GR and ER from implicit behavioral signals; (b) Apart from recognition experiments, we show that the employed devices capture meaningful perceptual information, as typified by gender and emotion-specific event related potentials (ERPs), and the extent of fixation over the eyes;  (c) We demonstrate a significant performance improvement in GR with deep learning and boosting methods over traditional approaches presented in \cite{maneesh2017acii,Bilalpur2017}; (d) The use of minimally intrusive, off-the-shelf and low-cost devices affirms the ecological validity of our study, and the utility of our experimental design for large-scale user profiling. 

Hereon, Section~\ref{RW} reviews related work. Section~\ref{MM} describes our study. Section~\ref{DA} examines \textit{explicit} user responses, which are correlated with \textit{implicit} behaviors in Section~\ref{IC}, followed by GR and ER experiments. Section~\ref{Con} summarizes and concludes the paper.

\section{Related Work}\label{RW}
Many works perform ER with \emph{implicit} behavioral signals such as eye movements, EEG, Electromyogram (EMG), Galvanic Skin Response (GSR)~\cite{zheng2014multimodal,liu2016multimodal,abadi2015decaf,Koelstra2012,subramanian2016ascertain}~\etc~However, very few works estimate \emph{soft biometrics} such as \emph{gender}, \emph{cognitive load} and \emph{personality traits} with such signals~\cite{wu2015human,Bilalpur18,Hoppe18}. Also, while some works isolate emotion and gender differences in eye movement and EEG responses to emotional faces~\cite{schurgin,zotto2015processing,katti2010making,Subramanian2014}, these differential features are never utilized for gender prediction. To position our work with respect to the literature, this section reviews related work on (a) user-centered ER, and (b) gender differences in emotional face processing.

\subsection{User-centered ER}
\begin{sloppypar}
Emotions evoked by multimedia stimuli are predicted via \textit{content-centered} or \textit{user-centered} methods. Content-centered methods attempt to find emotional multimodal features~\cite{Hanjalic2005,wang2006affective,vonikakis2017probabilistic,Shukla2017acm}, while user-centered methods monitor user behavioral cues (eye movements, EEG signals, \etc) to deduce the evoked emotion. As emotions are subjective, many user-centered approaches predict emotions by examining both \textit{explicit} and \textit{implicit} user behavioral cues. 
Conspicuous facial cues are studied to detect multimedia highlights in~\cite{joho2011looking}, while  physiological measurements are utilized to model emotions induced by music and movie scenes in~\cite{Koelstra2012,abadi2015decaf}. EEG and eye movements are two popular modalities employed for ER, and many works have used a combination of both~\cite{zheng2014multimodal,liu2016multimodal,maneesh2017acii} or either signal exclusively~\cite{li2015eeg,zheng2014eeg,Tavakoli15,subramanian2016ascertain,abadi2015decaf,katti2010making,Subramanian2014}. 
\end{sloppypar}

Valence (positive \textit{vs.} negative emotion) recognition from eye-gaze features is achieved in~\cite{Tavakoli15}. ER from EEG and pupillary responses is discussed in~\cite{zheng2014multimodal}. Deep unsupervised ER from raw EEG data is proposed in~\cite{li2015eeg}, and its effectiveness is shown to be comparable to designed features. Differential entropy EEG features are extracted to train an integrated deep belief network plus hidden Markov model for ER in~\cite{zheng2014eeg}. A differential auto-encoder that learns shared representations from EEG and eye-based features is proposed for valence recognition in~\cite{liu2016multimodal}. Almost all of these works employ lab-grade eye-trackers and EEG sensors which are bulky and intrusive, and therefore preclude naturalistic user behavior.


\subsection{Gender Differences in Emotion Recognition}
As facial emotions denote critical non-verbal communication cues in social interactions, many psychology studies have studied human FER. Certain facial features encode emotions better than others; the eyes, nose and mouth are the most \emph{attractive} facial regions~\cite{walker-smith,smith}. Visual attention is localized around the eyes for mildly emotive faces, but the nose and mouth attract substantial eye fixations in highly emotive faces~\cite{subramanian2011can}. An eye tracking study~\cite{schurgin} notes that distinct eye fixation patterns emerge for different facial emotions. The mouth is the most informative for the \textit{joy} and \textit{disgust} emotions, whereas eyes mainly encode information relating to \textit{sadness}, \textit{fear}, \textit{anger} and \textit{shame}. A similar study~\cite{aviezer} notes more fixations on the upper face half for \textit{anger} as compared to \textit{disgust}, while no differences are observed on lower face half for the two emotions. However, humans may find it difficult to distinguish between similar facial emotions-- examples are the high overlap rate between the \textit{fear}--\textit{surprise} and \textit{anger}--\textit{disgust} emotion pairs~\cite{smith,etcoff}.

Multiple works have discovered gender differences during facial emotion processing. Females are generally better at FER irrespective of age~\cite{sullivan}. Other FER studies~\cite{bassili,montagne,halljudith} also note that females recognize facial emotions more accurately than males, even under partial information. Some evidence also points to females achieving faster FER than males~\cite{halljessica,rahman}. Gender differences in gaze patterns and neural activations have been found while viewing emotional faces; female tendency to fixate on the eyes positively correlates with their ER capabilities, while men tend to look at the mouth for emotional cues~\cite{sullivan,halljessica}. Likewise, EEG Event-related potentials (ERPs) reveal that negative facial emotions are processed differently and rapidly by women, and do not necessarily entail selective attention towards emotional cues~\cite{Lithari2010,zotto2015processing}. 

An exhaustive review of GR methodologies is presented in~\cite{wu2015human}, and the authors evaluate GR methods using metrics like \textit{universality, distinctiveness, permanence} and \textit{collectability}. While crediting bio-signals like EEG and Electrocardiography (ECG) for their accuracy and trustworthiness, authors also highlight the invasiveness of bio-sensors. The sensors used in this work are minimally intrusive, enabling naturalistic user experience, while also recording meaningful emotion and gender-related information. Among user-centric GR works, EEG and speech features are proposed for age and gender recognition in~\cite{nguyen2013age}.

\subsection{Analysis of Related Work}      
Close examination of the literature reveals (1) Many works achieve ER from user-centered cues, both conspicuous and latent, and a handful have discovered gender differences in gaze patterns and neural activations; nevertheless, \emph{very few works expressly predict gender from implicit user cues}. Differently, we employ implicit signals for GR and ER, and achieve reliable gender and valence detection (AUC $>$ 0.9); (2) Our GR/ER features are acquired from low cost, off-the-shelf sensors, which record inferior user signals even while enabling natural user behavior. We nevertheless show how these sensors capture meaningful information via the analysis of ERPs and fixation distribution patterns; (3) Different to prior works which either analyze \emph{explicit} or \emph{implicit} user responses to discover gender differences, or do not expressly isolate bio-signal patterns; in contrast, we show multiple similarities among explicit and implicit user behaviors to validate our findings. 
\section{Materials and Methods}\label{MM}
Our study objective was to examine user behavior while viewing unoccluded/partly occluded emotional faces, and predict \emph{user gender} therefrom. We hypothesized that gender differences would be captured via EEG and eye-gaze patterns. Also, eye movements are known to be characteristic of the perceived facial emotion~\cite{schurgin}, which enables inference of the {stimulus emotion}; we limit ourselves to predicting the \emph{stimulus valence}, \ie, whether the face presented to the viewer exhibits a positive or negative emotion?

\begin{figure*}[t]
	\includegraphics[width=0.95\linewidth]{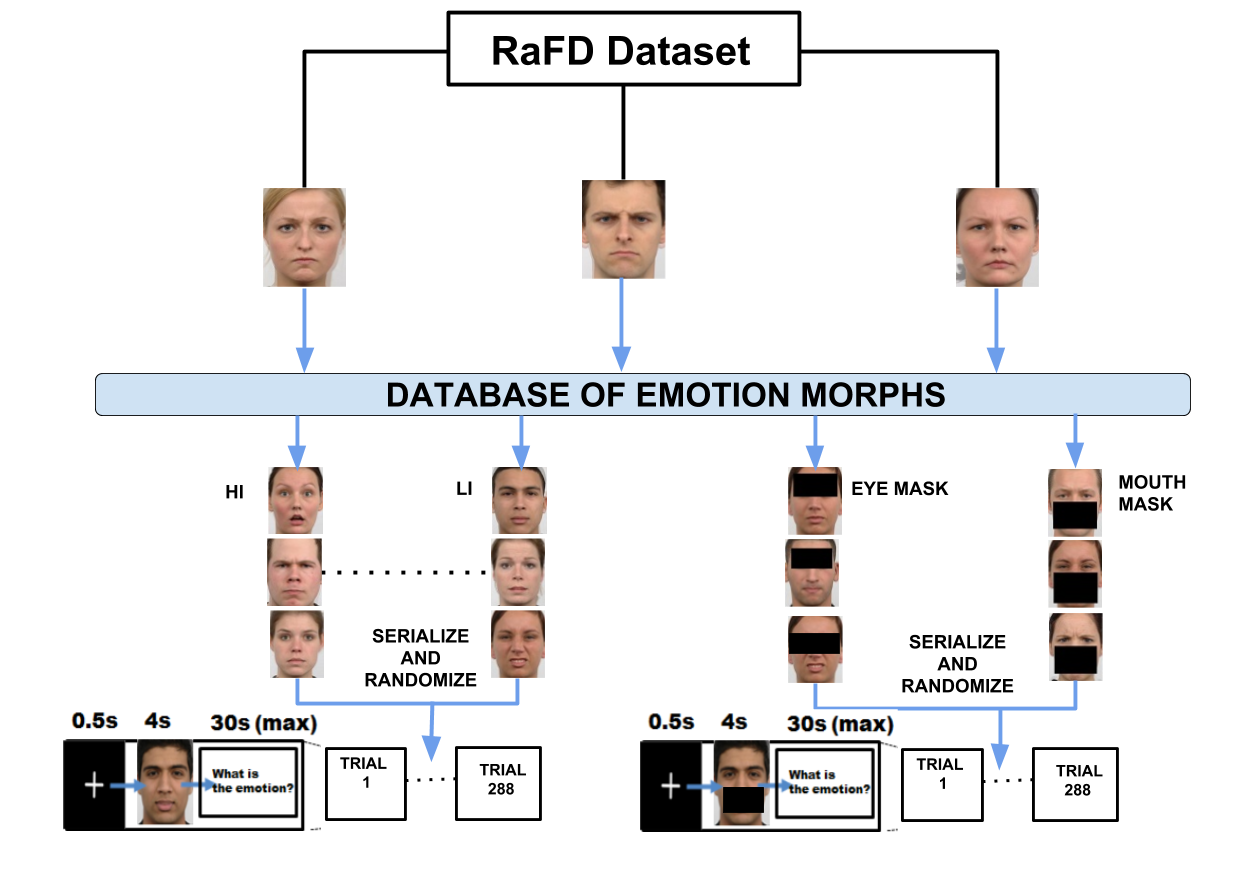} \vspace{-0.5cm}
    \caption{\label{proto} \textbf{Experimental Protocol:} Viewers were required to recognize the facial emotion from either an \textit{unmasked} face (Session 1), or from an \textit{eye}/\textit{mouth} masked face (Session 2). Trial timelines for the two conditions are shown at the bottom.} \vspace{-0.5cm}
\end{figure*}

We designed a study to examine gender differences in visual emotional face processing with a) fully visible faces, and (b) faces with the \textit{eye} and \textit{mouth} regions occluded via a rectangular mask (Fig.~\ref{proto}). Salience of the occluded features towards conveying facial emotions is well known~\cite{subramanian2011can,schurgin}. Specifically, we considered emotional faces corresponding to four conditions: exhibiting \textit{high} intensity (\textbf{HI}) and \textit{low} intensity (\textbf{LI}) emotions, and additionally, \textit{high} intensity emotions upon occluding the \textit{eye} ({\textbf{eye-mask}}) or \textit{mouth} ({\textbf{mouth-mask}}) regions (Fig.~\ref{proto}). Since we hypothesized that emotion perception under facial occlusion would be considerably difficult (cf. Table~\ref{tab:GenResults1}), we did not study masked and mildly emotive faces. \\

\noindent \textbf{Participants:} 28 students from different nationalities (14 male, age $26.1\pm7.3$ and 14 female, age $25.5\pm6$), with normal or corrected vision, took part in our study. All users provided informed consent, and were presented a token fee for participation as directed by the ethics committee.  \\

\noindent \textbf{Stimuli:} We used emotional faces of 24 models (12 male, 12 female) from the {Radboud Faces Database} (RaFD)~\cite{Rafd}. RaFD includes facial emotions of 49 models rated for \textit{clarity}, \textit{genuineness} and \textit{intensity}, and the 24 models were chosen such that their Ekman facial emotions (\textbf{A}nger, \textbf{D}isgust, \textbf{F}ear, \textbf{H}appy, \textbf{Sa}d and \textbf{Su}rprise) were roughly matched based on these ratings. We then \textit{morphed} the emotive faces from \textit{neutral} (0\% intensity) to \textit{maximum} (100\% intensity) to generate intermediate morphs in steps of 5\%. Derived morphs with 55--100\% intensity were used as \textbf{HI} emotions, and 25--50\% were used as \textbf{LI} emotions. \textit{Eye} and \textit{mouth-masked} faces were automatically generated upon locating facial landmarks via Openface~\cite{Baltrusaitis2016} over the HI morphs. The \textit{eye mask} covered the eyes and nasion, while the \textit{mouth mask} covered the mouth and the nose ridge. All stimuli were resized to 361$\times$451 pixels, encompassing a visual angle of 9.1$^\circ$ and 11.4$^\circ$ about $x$ and $y$ at 60cm screen distance. \\

%
\noindent \textbf{Protocol:} The experimental protocol is outlined in Fig.\ref{proto}, and 
involved the presentation of \textit{unmasked} and \textit{masked} faces to viewers over two separate sessions, with a break in-between to avoid fatigue. We chose one face per model and emotion, resulting in 144 face images (1 morph/emotion $\times$ 6 emotions $\times$ 24 models). In the first session (\textit{no-mask} condition), these faces were shown in random order and were again re-presented randomly with an \textit{eye} or \textit{mouth mask} in the second session. We ensured a 50\% split of the HI and LI morphs in the first session, and eye/mouth-masked faces in the second.

During each trial, an emotional face was displayed for 4s preceded by a fixation cross for 500 ms. The viewer then had a maximum of 30s to make {one} out of {seven} choices concerning the facial emotion (six Ekman emotions plus neutral) via a radio button. Neutral faces were only utilized for morphing purposes and not used in the experiment. Viewers' EEG signals were acquired via the 14-channel \textit{Emotiv Epoc+} device, and eye movements were recorded with the \textit{Eyetribe} tracker during the trials. The face-viewing experiment was split into 4 segments to facilitate sensor re-calibration and minimize data recording errors, and took about 90 minutes to complete. 

\section{User Responses}\label{DA}
We first compare male and female sensitivity to emotions based on \textit{explicitly} observed user response times (RTs) and recognition rates (RRs), and will then proceed to examine their \textit{implicit} eye movement and EEG responses. Our experimental design involved four \textit{stimulus types} (HI, LI, eye and mouth mask), and two \textit{user types} (male and female), resulting in $4 \times 2$ factor conditions.


\begin{figure}[t]
\includegraphics[width=0.95\linewidth]{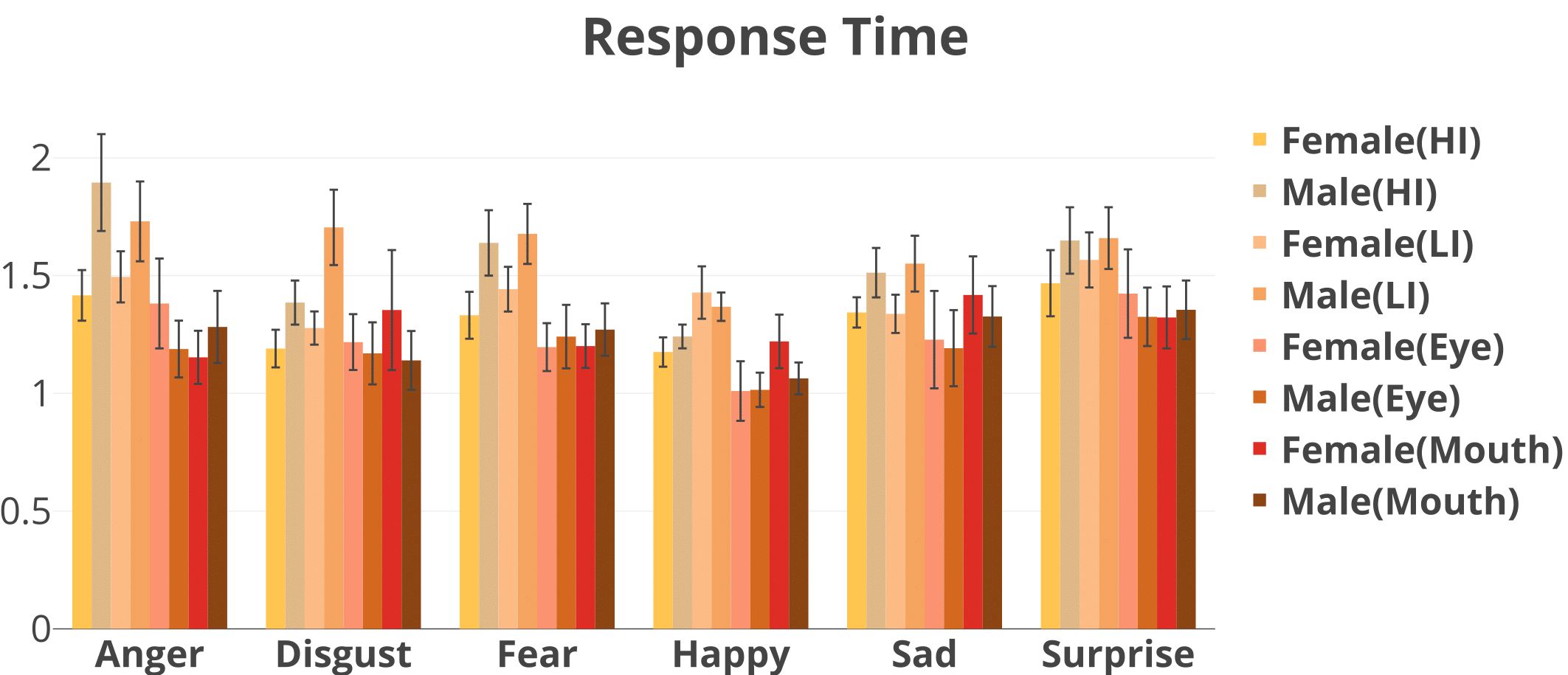}
\caption{\label{RT_HR} \textbf{Emotion-wise RTs} of females and males in the various conditions. $y$-axis denotes response time in seconds. Error bars denote unit standard error (best-viewed in color).}
\vspace{-0.4cm}
\end{figure}

\begin{figure}[t]
\centering
\includegraphics[width=0.95\linewidth]{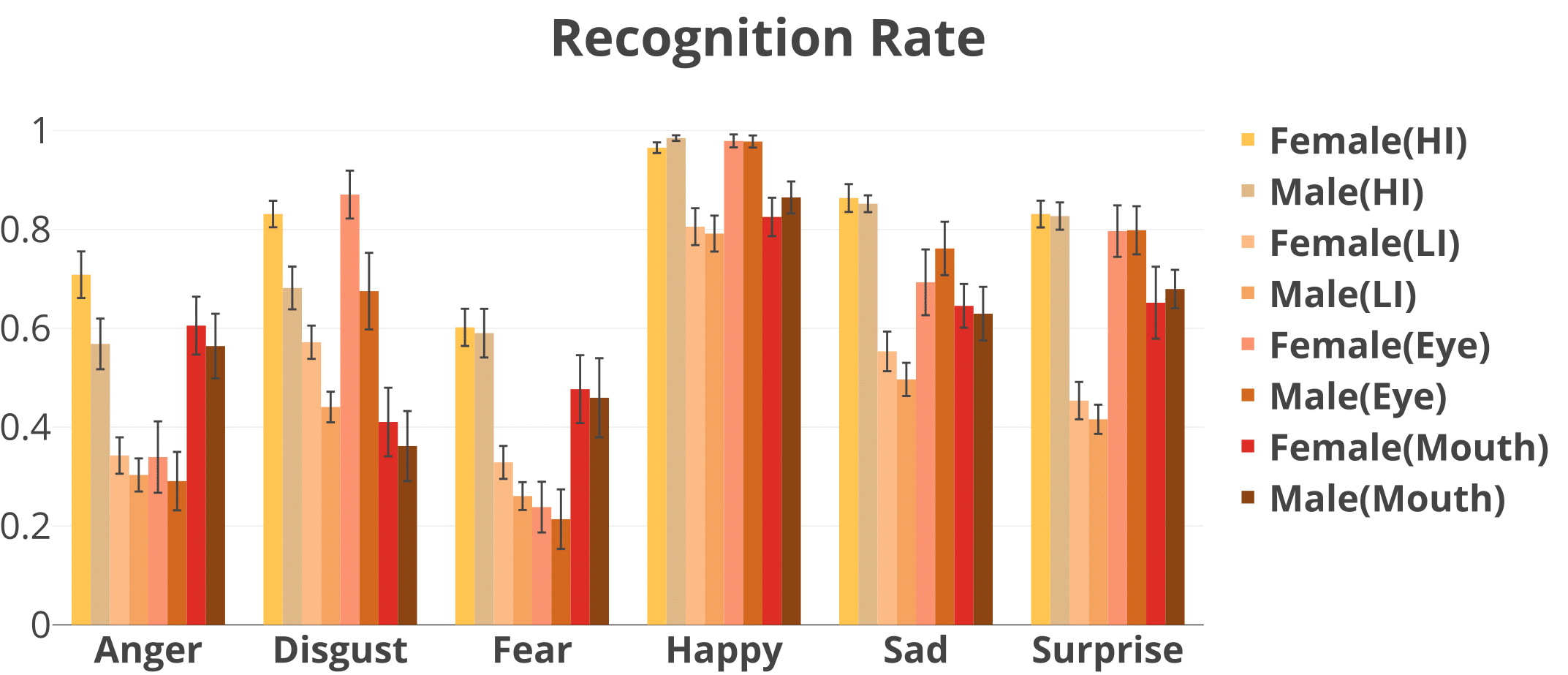}\vspace{-.2cm}
\caption{\label{RRnfix} \textbf{Emotion-wise RRs} in various conditions.} \vspace{-.5cm}
\end{figure}

%



\subsection{Response time (RT)}\label{RT_mask}
Overall user RTs for the HI, LI, {eye mask} and {mouth mask} conditions were respectively found to be 1.44 $\pm$ 0.24, 1.52 $\pm$ 0.05, 1.17 $\pm$ 0.12 and 1.25 $\pm$ 0.09 seconds, implying that FER was fairly instantaneous, and viewer responses were surprisingly faster with masked faces. Fine-grained comparison of male (\textit{m}) and female (\textit{f}) RTs across stimulus types (Fig.~\ref{RT_HR}) revealed that females ($\mu_{\text{RT}} = 1.40 \pm 0.10$s) were generally faster than males ($\mu_{\text{RT}} = 1.60 \pm 0.10$s) at recognizing {HI} emotions. There was no significant difference in RTs for {LI} emotions. Female alacrity nevertheless decreased for masked faces, with males responding marginally faster for \textit{eye masked} ($\mu_{\text{RT}}\text{(\textit{m})} = 1.13 \pm 0.11$s vs $\mu_{\text{RT}}\text{(\textit{f})} = 1.21 \pm 0.13$s), and both genders responding with similar speed for \textit{mouth masked} faces ($\mu_{\text{RT}}\text{(\textit{m})} = 1.24 \pm 0.10$s vs $\mu_{\text{RT}}\text{(\textit{f})} = 1.25 \pm 0.09$s). \\

\subsection{Recognition Rates}\label{RR_mask}
While females recognized facial emotions marginally faster, we examined if they also achieved superior FER. Overall, RRs for unoccluded {HI} emotions ($\mu_{\text{RR}}=77.6$) were expectedly higher than for \textit{eye-masked} ($\mu_{\text{RR}}=59.7$), \textit{mouth-masked} ($\mu_{\text{RR}}=63.5$) and {LI} emotions ($\mu_{\text{RR}}=49.1$). Happy faces were recognized most accurately in all four conditions. Specifically focusing on gender differences (Fig.~\ref{RRnfix}), females recognized facial emotions more accurately than males and this was particularly true for \textit{negative} (A, D, F, S) emotions; male vs female RRs for these emotions differed significantly in the HI ($\mu_{\text{RR}}{(m)} =54.3$ vs $\mu_{\text{RR}} {(f)} =61.2, p<0.05$) and \textit{eye mask} conditions ($\mu_{\text{RR}}{(m)} =51.8$ vs $\mu_{\text{RR}}{(f)} =58.1, p<0.05$), and marginally for the \textit{mouth mask} condition 
($\mu_{\text{RR}}\text{(m)} =52$ vs $\mu_{\text{RR}} \text{(f)} =55.8, p=0.08$). Males marginally outperformed females in the  LI condition ($\mu_{\text{RR}}\text{(m)} = 49.8$ vs $\mu_{\text{RR}} \text{(f)} =47.7$). Overall, (a) HI emotion morphs were recognized more accurately than LI morphs, (b) females recognized negative emotions better, (c) \textit{Happy} was the easiest emotion to recognize, and (d) Higher RRs (across gender) were noted in the \emph{eye-mask} condition for four of the six Ekman emotions, implying that deformations around the mouth were more informative for FER under occlusion. 

\section{Analyzing Implicit Responses}\label{IC}
As females achieved quicker and superior FER for negative emotions, we hypothesized that these behavioral differences should also reflect via \emph{implicit} eye gaze and EEG patterns. We first describe the EEG and eye-movement descriptors employed for analyses, before discussing (stimulus) emotion and (user) gender recognition results.

\subsection{EEG preprocessing}\label{preprocessing}
We extracted EEG epochs for each trial (4.5s of stimulus-plus-fixation viewing time at 128 Hz sampling rate), and the 64 leading pre-stimulus samples were used to remove DC offset. This was followed by (a) EEG band-limiting to within 0.1--45 Hz, (b) Removal of noisy epochs via visual inspection, and (c) Independent component analysis (ICA) to remove artifacts relating to eye-blinks, and eye/muscle movements. Muscle movement artifacts in EEG are mainly concentrated in the 40--100 Hz band, and are removed upon band-limiting and via inspection of ICA components. Finally, a 7168 dimensional (14 channel $\times$4s$\times$128 Hz) EEG feature vector was generated and fed to different classifiers for GR and ER (Section~\ref{experiments}). 

\subsubsection{Event Related Potentials}\label{ERP}   
Event Related Potentials are time-locked neural responses related to sensory and cognitive events, and denote the EEG response averaged over multiple users and trials. As examples, P300 and N100, N400 are exemplar ERPs which are typically noted around 300, 100 and 400 ms post stimulus onset. ERPs occurring within 100 ms post stimulus onset are \emph{stimulus-related} (exogenous), while later ERPs are \emph{cognition-related} (endogenous). We examined the leading 128 EEG epoch samples (one second of data) for ERP patterns relating to emotion and gender. 

\begin{figure*}[t]
\includegraphics[width=0.48\linewidth]{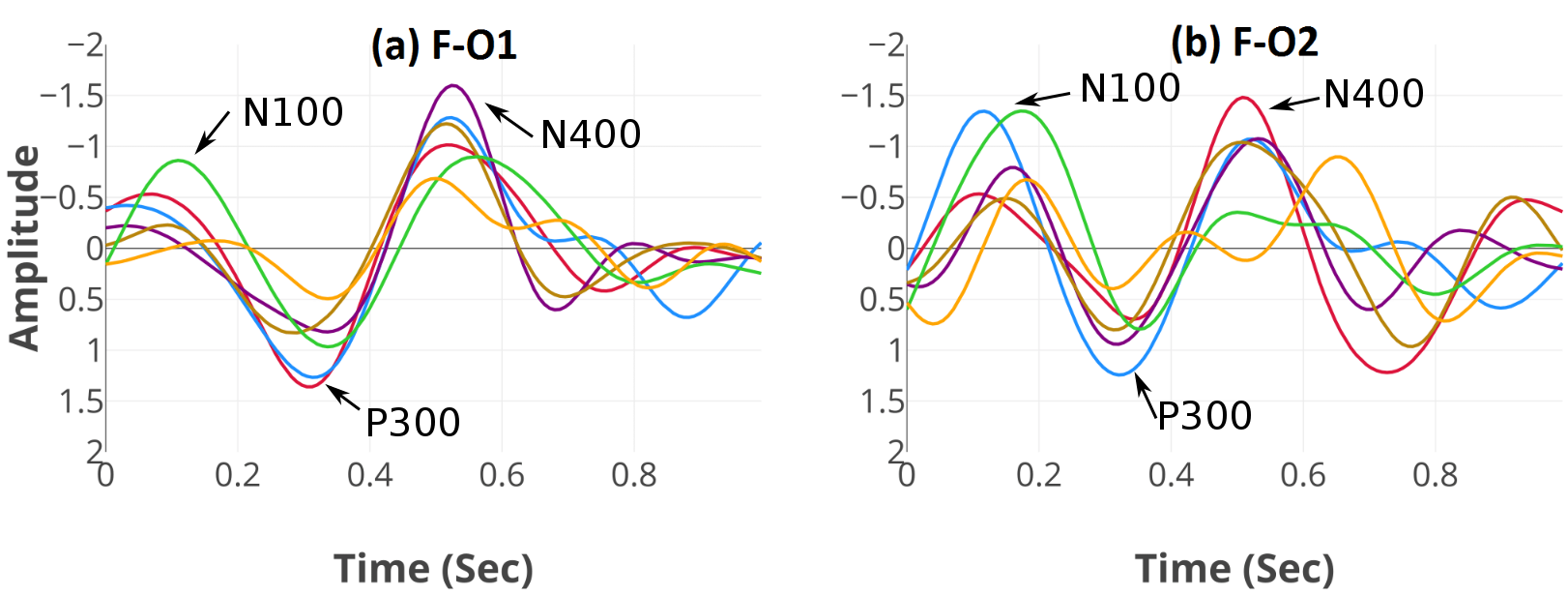}
\hspace{-1cm}
\includegraphics[width=0.48\linewidth]{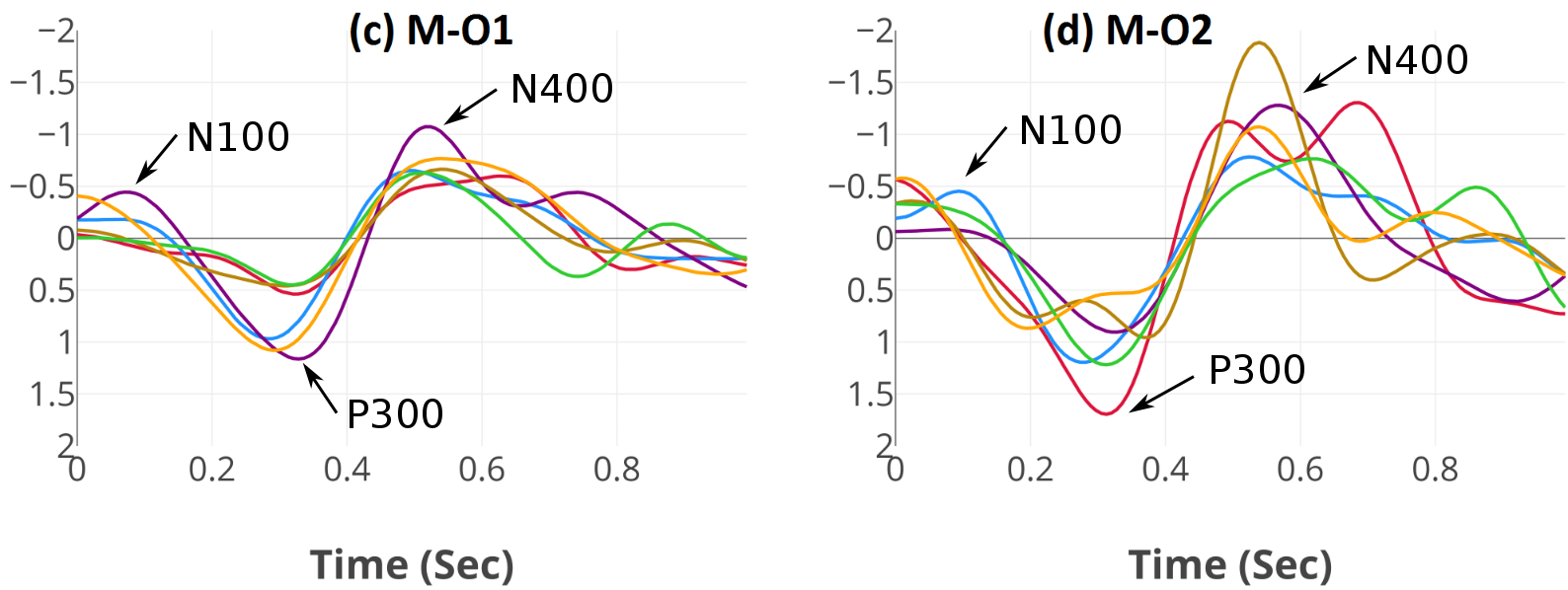}
\vspace{-2cm}
\includegraphics[width=0.48\linewidth]{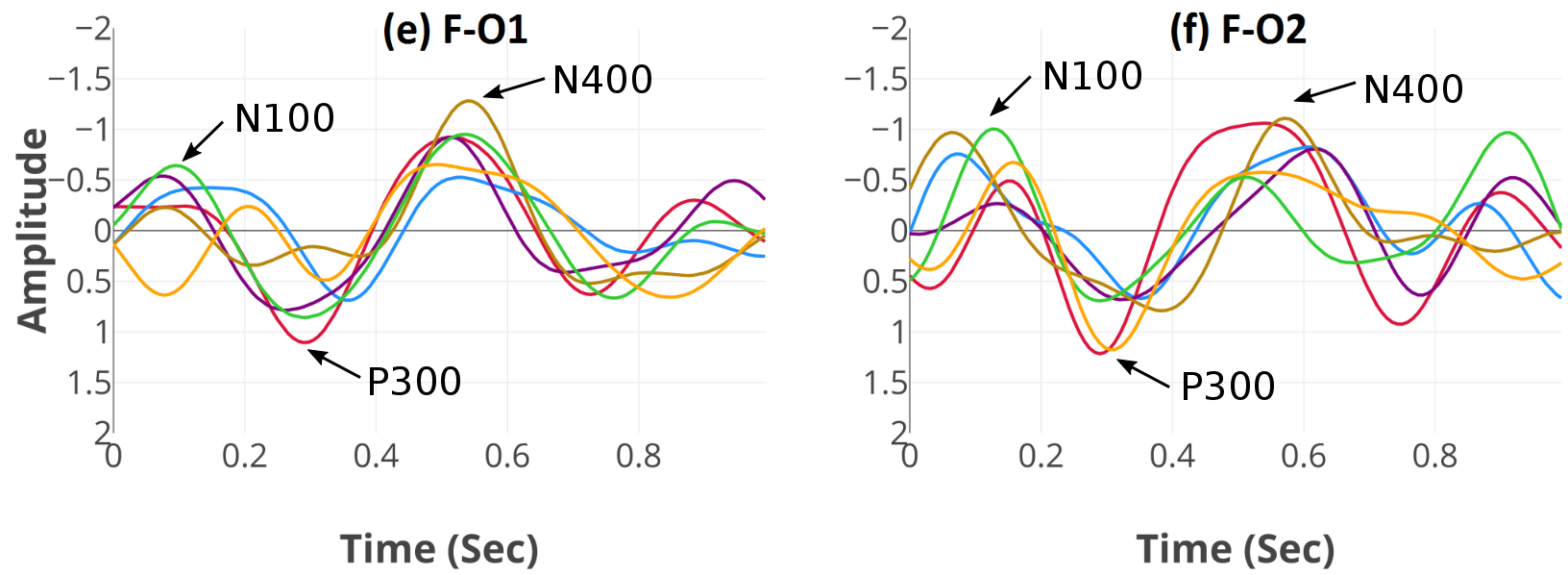}
\includegraphics[width=0.48\linewidth]{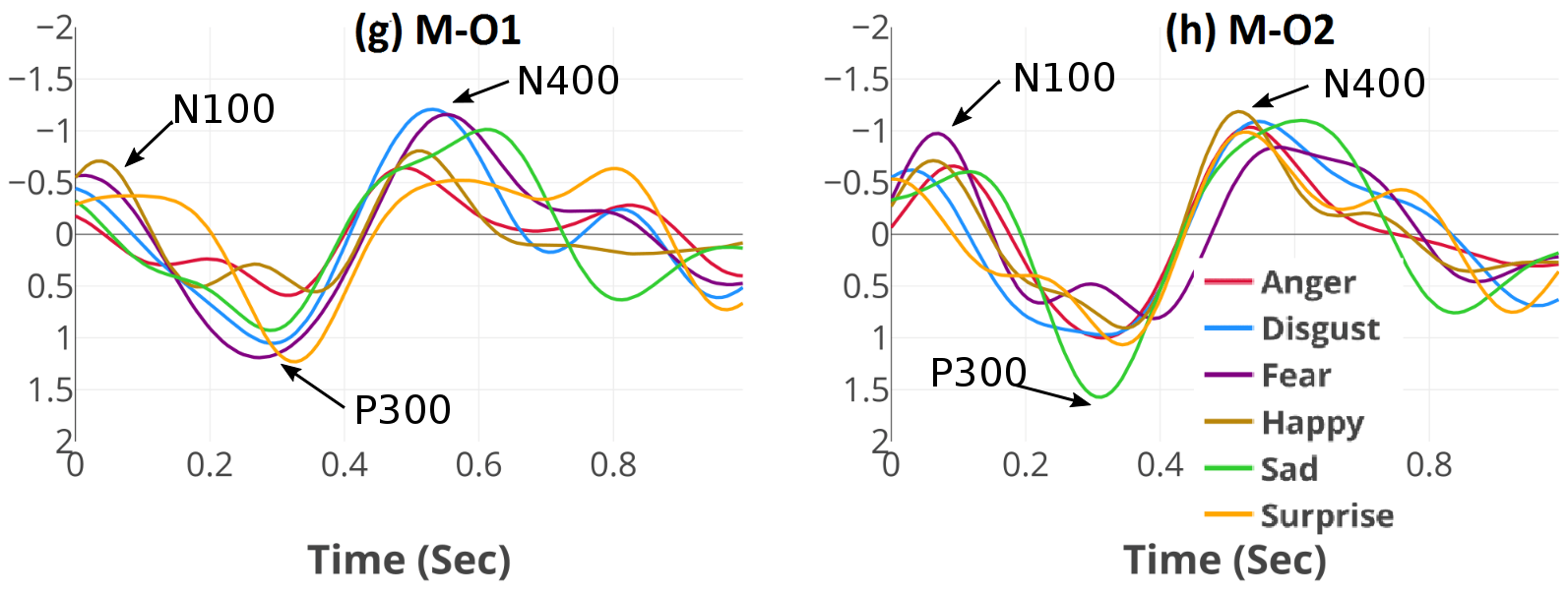}
\vspace{1.4cm}
\caption{\label{ERP_HL} \textbf{ERPs} for {HI}  morphs (top) and {LI} morphs (bottom): (left to right) O1 and O2 ERPs for females and males. $y$-axis shows ERP amplitude in ${{\mu V}}$, refer (h) for legend. As per convention, ERPs are plotted upside down (view in color).} \vspace{-.4cm}
\end{figure*}

\begin{figure}[t]
    \centering
	  \includegraphics[width = 1.0\linewidth]{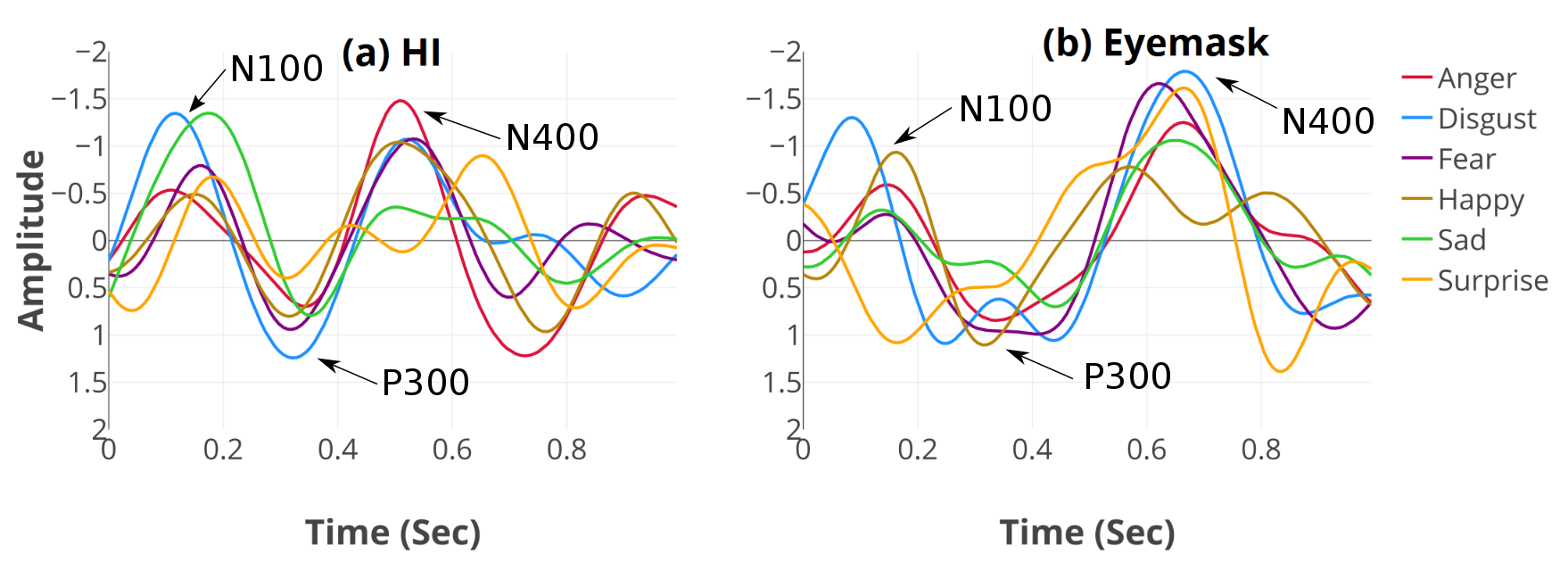} \vspace{-0.3cm}
    \caption{\textbf{Female ERPs} in the (left) {HI}  and (right) {eye mask} conditions from the O2 channel. Best viewed in color.} \vspace{-0.5cm}
    \label{ERPs1}
\end{figure}

Prior works have observed ERP-based gender differences from lab-grade sensor recordings~\cite{Lithari2010,Muhl14,zotto2015processing}. Specifically,~\cite{Lithari2010} notes enhanced negative ERPs for females in response to negative valence stimuli. However, capturing ERPs with commercial devices is challenging due to their low signal-to-noise ratio~\cite{Vi2014}. Figs.~\ref{ERP_HL} and~\ref{ERPs1} present the P300, visual N100 and N400 ERP components in the occipital O1 and/or O2 electrodes (see Fig.~\ref{Emotiv} for sensor positions) corresponding to various face morphs. Note that the occipital lobe is the \textit{visual processing center} in the brain, as it contains the primary visual cortex. 

Comparing O1/O2 male and female ERPs for positive (H, Su) and negative (A, D, F, Sa) emotions, no significant differences can be observed between male positive and negative ERP peaks for HI or LI faces (columns 3,4 in Fig.\ref{ERP_HL}). However, we observe stronger N100 and P300 peaks in the negative female ERPs for both HI and LI faces (columns 1,2). Also, a stronger female N400 peak can be noted for HI faces consistent with prior findings~\cite{Lithari2010}. Contrastingly, lower male N100 and P300 latencies are observed for positive HI emotions, with the pattern being more obvious at O2. Likewise, lower male N400 latencies can be generally noted at O2 for positive emotions. The positive vs negative ERP difference for females is narrower for LI faces, revealing the difficulty in identifying mild LI emotions. That LI faces produce weaker ERPs at O1 and O2 than HI faces further supports this observation. 

Fig.~\ref{ERPs1} shows female ERPs observed in the occipital O2 electrode for the HI and \textit{eye mask} conditions. Clearly, one can note enhanced N100 and P300 ERP components for negative HI emotions (Fig.~\ref{ERPs1}(left)). This effect is attenuated in the \textit{eye mask} (Fig.~\ref{ERPs1}(right)) and \textit{mouth mask} cases, although one can note stronger N400 amplitudes for D and F with \textit{eye mask}. This ERP pattern is invisible for males, confirming thir gender-specificity. Overall, ERP patterns affirm that gender differences in emotional face processing can be reliably isolated with the low-cost \textit{Emotiv} device. 

\subsection{Eye-tracking analysis}
Gender differences in gaze patterns during emotional face processing have been noted by prior works~\cite{schurgin,sullivan}. We used the low-cost \textit{Eyetribe} device with 30 Hz sampling to record eye movements. Raw gaze data output by the tracker were processed to compute \textit{fixations} (prolonged gazing at scene regions to assimilate visual information) and \emph{saccades} (transition from one fixation to another) via the EyeMMV toolbox~\cite{krassanakis2014eyemmv}. Upon extracting fixations and saccades, we extracted features employed for valence recognition in~\cite{Tavakoli15} to compute an 825-dimensional feature vector for our analyses.

\begin{figure*}[t]
\centering
\includegraphics[width = 0.95\linewidth]{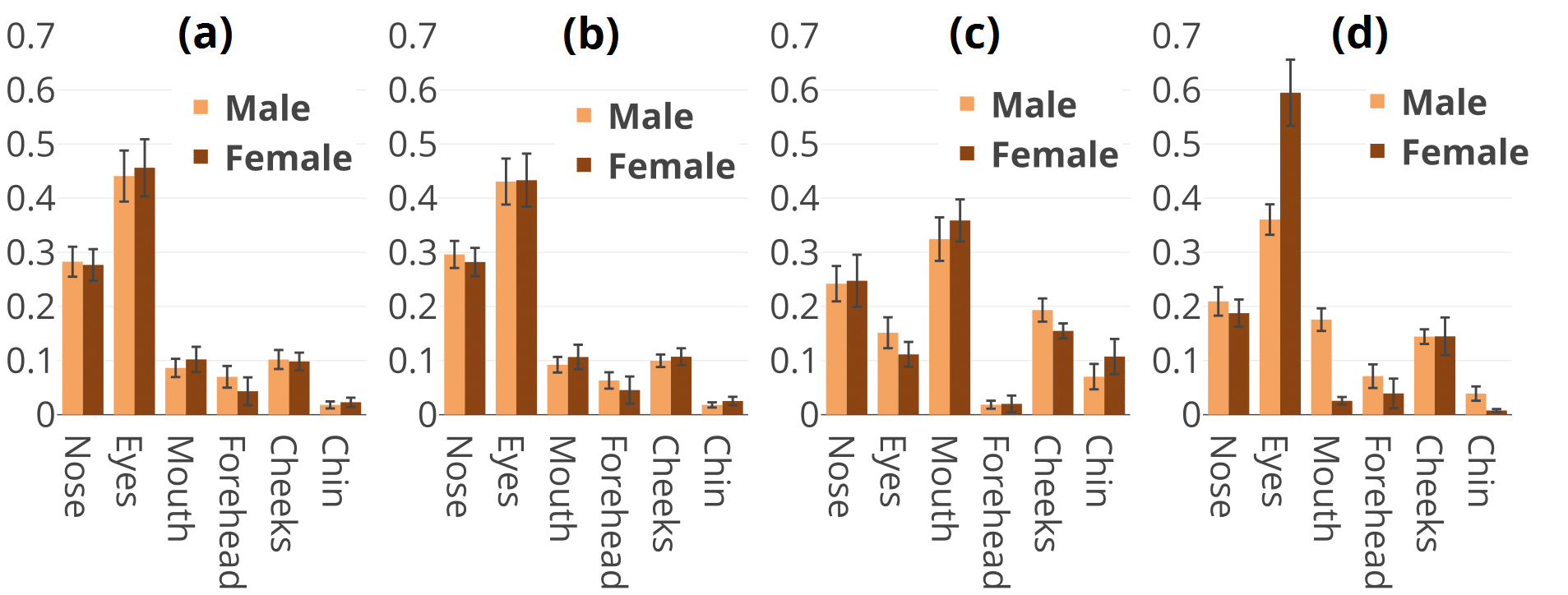}
\caption{(a--d) \textbf{Fixation duration} distributions for males and females in the {HI}, {LI}, {eye} and {mouth-masked} conditions.}\vspace{-0.5cm}
\label{Fix_dur_dist}
\end{figure*}

\subsubsection{Fixation analysis}\label{Fix_anal}
To study gender differences in fixating patterns, we computed the distribution of fixation duration (FD) over six facial regions, namely, \textit{eyes}, \textit{nose}, \textit{mouth}, \textit{cheeks}, \textit{forehead} and \textit{chin}. Fig.~\ref{Fix_dur_dist} presents the males and female FD distribution over these facial regions for various conditions. For both genders, the time spent examining the eyes, nose and mouth accounted for over 80\% of the total FD, with eyes ($\approx$45\%) and nose ($\approx$30\%) attracting maximum attention as observed in \cite{schurgin,katti2010making,subramanian2011can}. Relatively similar FD distributions were noted for both genders with HI and LI morphs (Fig.~\ref{Fix_dur_dist} a,b). 


Fig.\ref{Fix_dur_dist}(c,d) present FD distributions in the \textit{eye} and \textit{mouth mask} conditions. An independent $t$-test revealed a significant difference (\textit{p}$<$0.05) between male and female FDs for the eye region in the \textit{mouth mask} condition; prior works~\cite{wells2016identification} have observed that females primarily look at the eyes for emotional cues, which is mirrored by longer FDs around the eyes in the HI and \textit{mouth mask} conditions. Fig.\ref{Fix_dur_dist}(c) shows that when eye information is unavailable, females tend to focus on the mouth and nasal regions for FER.

\subsection{Experiments and Results}
\label{experiments}
Having noticed gender-specific patterns in user responses, EEG ERPs and eye movements, we attempted binary emotion recognition (ER) and gender recognition (GR) employing EEG features, eye-based features and their combination for the various conditions. {Recognition was attempted only on trials where viewers correctly recognized the presented facial emotion.}
We considered (i) EEG features, (ii) eye-based features, (iii) concatenation of the two (\textit{early fusion} or EF), and (iv) probabilistic fusion of the EEG and eye-based classifier outputs (\textit{late fusion} or LF) for our analyses. The $W_{est}$ technique~\cite{koelstra2012fusion} was used to fuse the EEG and eye-based outputs, and denotes \textit{maximum possible} recognition performance as optimal weights maximizing the test AUC metric were determined via a 2D grid search.

We considered the area under ROC curve ({\textbf{AUC}}) plotting true vs false positive rates, as the performance metric for benchmarking. AUC is suitable for evaluating classifier performance on unbalanced data, and a random classifier will achieve an AUC of 0.5.  As we attempted recognition with few training data, we report ER/GR results over five repetitions of 10-fold cross validation (CV) (\ie, total of 50 runs). CV is typically used to overcome the \textit{overfitting} problem, and train generalizable classifiers on small datasets. 

\subsubsection{Baseline Classification Approaches} 
As baselines, we considered the Naive-Bayes (NB), linear SVM (LSVM) and radial-basis SVM (RSVM) classifiers. NB is a generative classifier that estimates the test label based on the \emph{maximum a-posteriori} criterion,  $p(C\mid X)$, assuming class-conditional feature independence. $C$ and $X$ respectively denote test class-label and feature vector. LSVM and RSVM denote the linear and radial basis kernel versions of the Support Vector Machine (SVM) classifier. SVM hyperparameters $C$ (LSVM) and $\gamma$ (RSVM) were tuned from within $[10^{-4},10^{4}]$ via an inner 10-fold CV on the training set. 


\subsubsection{CNN for EEG-based GR \& ER} 
Deep learning frameworks have recently become popular due to their ability to automatically learn optimal task features from raw data, thereby obviating the need for data cleaning and feature extraction. However, unlike in  image or video processing, temporal dynamics of the human brain are largely unclear; designing relevant features is therefore hard. We hypothesized that CNNs would encode EEG patterns efficiently, and also be robust to artifacts. These factors inspired us to feed \textit{raw} EEG data to a CNN, without any data processing (as in Sec.~\ref{preprocessing}); preprocessed EEG data was nevertheless fed to the above baseline classifiers.


\begin{figure*}
\vspace{0.05cm}
\centering
\includegraphics[width = 0.9\linewidth]{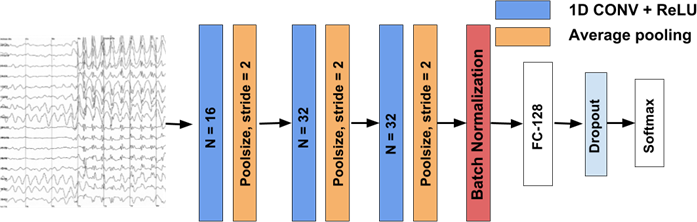}
\caption{\textbf{CNN architecture} showing various layers in the model and parameters.}\vspace{-0.5cm}
\label{DLarch}
\end{figure*}

We adopted a 3-layer Convolutional Neural Network (CNN)~\cite{MOHAMMADIANRAD2018180} to learn a robust EEG representation for gender and valence recognition. Three convolution layers together with rectified linear unit (ReLU) activation, and average pooling layers are stacked to learn EEG descriptors (Fig.~\ref{DLarch}). The convolutions employed are 1-dimensional along time. Batch normalization~\cite{pmlr-v37-ioffe15} is used after the third CNN layer to minimize covariate shift and accelerate training. To prevent overfitting, we used dropout after the fully connected layer with 128 neurons. A {softmax} over two output neurons was used for classification.

The number of kernels increase with network depth (cf. Fig~\ref{DLarch}) analogous to the VGG architecture~\cite{Simonyan14c}. We optimized the network for categorical cross-entropy loss using stochastic gradient descent with Nesterov momentum and weight decay. To decrease data dimensionality and avoid overfitting without sacrificing temporal EEG dependencies, we downsampled EEG data to 32Hz (reducing per-channel EEG dimensionality to 32$\times$4=128). A tenth of training data was used for validation and early-stopping enforced to prevent overfitting. CNN hyperparameters specified in Table~\ref{network_hyp} were either adopted from~\cite{MOHAMMADIANRAD2018180} or upon cross-validation. 

\begin{table}[]
\fontsize{6.5}{6.5}\selectfont
\centering
\caption{Summary of the \textbf{deep CNN model}.}
\vspace{-.2cm}
\scalebox{1.3}{%
\label{network_hyp}
\begin{tabular}{ll}
\hline
\hline
Parameter        & Value    \\ \hline
Learning rate          & 0.01     \\
Kernel shape           & 3        \\
Stride size            & 2        \\
Pool size              & 2        \\
Batch size             & 32       \\
\# Kernels(layer wise) & 16,32,32 \\
Momentum			   & 0.9	  \\
Weight decay		   & 0.0001   \\
Dropout                & 0.1     \\
\hline
\hline
\end{tabular}}
\vspace{-.5cm}
\end{table}

The CNN was traditionally trained for GR with \textit{He normal} initializers (normally distributed weights). Conversely, we adopted two-stage training for ER. In the first stage, the model was pre-trained with EEG data acquired over all four experimental conditions (Fig~\ref{proto}); the second stage involved fine-tuning with data for a specific condition. The objective here was to extract valence-related features irrespective of stimulus type in the first stage, and fine-tune them to learn condition-specific descriptors. Experiments were designed using Keras~\cite{chollet2015keras} with Tensorflow back-end on a 16 GB CPU machine and an Intel i5 processor.
%
\subsubsection{AdaBoost for ER from eye-gaze} 
We employed the Adaboost classifier for ER from eye-gaze features. AdaBoost is an ensemble classifier popularly used for face detection~\cite{Viola:2004:RRF:966432.966458} and FER~\cite{silapachote2005feature}. Adaboost combines a number of weak classifiers (decision stumps), each marginally better than random, to cumulatively achieve optimal classification. The class label is based on the weighted output of each weak classifier. Our features (Sec.~\ref{preprocessing}) and classifier design were inspired by~\cite{silapachote2005feature}, who capture local dependencies via histograms of gaze measures like fixations, saccades and saliency. We employed SAMME (Stagewise Additive Modeling using a Multi-class Exponential loss function)~\cite{hastie2009multi} for training.  Similar to~\cite{silapachote2005feature}, local Gaussian and Gabor features were extracted at multiple scales to generate composite features for AdaBoost.  

We only obtained sparse gaze features in our study as viewer gaze was (a) localized to face regions of emotional importance and (b) recorded via a low-cost and low sampling rate device. We firstly performed feature selection to prune features relevant for ER. Feature selection was based on sequential addition (or removal) for optimal performance~\cite{Tavakoli15} using sequential forward (SFS) or backward (SBS) selection. User data compiled for this study along with the CNN and Adaboost models employed for GR and ER respectively are available at \url{https://github.com/bmaneesh/emotion-xavier}.

\subsection{Results} \label{Results}
\subsubsection{Emotion Recognition} \label{ER}
\begin{figure}[t]
\includegraphics[width=0.95\linewidth]{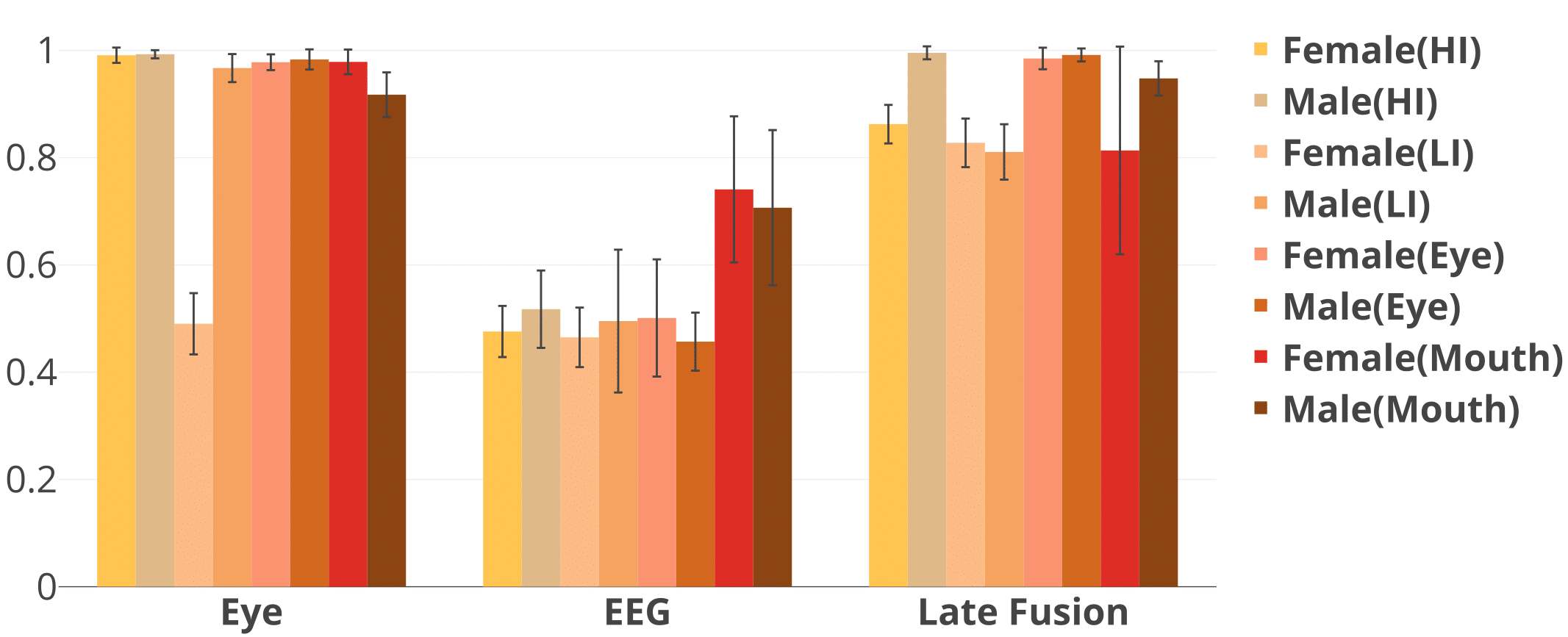}
\vspace{-0.1cm}
\caption{\textbf{Valence recognition} results for different conditions.
} 
\label{Rec_Emotiv} \vspace{-.5cm} 
\end{figure}

We modeled stimulus ER as a binary classification problem, where the objective is to categorize \textit{positive} (H, Su) and \textit{negative} (A, D, F, Sa) valence stimuli employing gaze and EEG-based cues. ER results with (CNN-based) EEG, (Adaboost-based) eye gaze features, and late fusion of two modalities, with data acquired for different conditions characterized by user-gender (M/F) and stimulus type (HI/LI/eye-mask/mouth-mask) are shown in Fig.~\ref{Rec_Emotiv}. Evidently, gaze features perform significantly better than EEG. Gaze features achieve near-ceiling valence recognition barring the case where female users view LI emotions. Optimal performance is noted with male data for HI emotions (AUC = 0.99) similar to ~\cite{maneesh2017icmi}; ER from female eye-gaze data on HI emotions (AUC = 0.98) is also high.
 
On the other hand, ER results with EEG largely produced near-chance performance, with only male and female data for the {mouth-masked} condition being exceptions. These results reveal that the raw EEG features are not optimal for ER. Given the vast difference in ER performance with the gaze and EEG modalities, late fusion results are only occasionally superior to unimodal ones. Fusion slightly outperforms unimodal methods for the {eye mask} condition for both males (0.98 vs. 0.99) and females (0.97 vs. 0.98), and noticeably with males for {mouth mask} faces (0.91 vs. 0.94).

Overall, while recognizing the stimulus facial emotion was not our primary objective, the extracted gaze features are nevertheless highly effective for valence recognition (VR).  Adaboost employing gaze features significantly outperforms the CNN trained with EEG data. The ability of eye movements to characterize emotional differences is unsurprising. Distinctive eye movement patterns have been observed while scanning scenes and faces with different emotions~\cite{schurgin,Subramanian2014,Tavakoli15}. Eye-based features are found to achieve 52.5\% VR accuracy in~\cite{Tavakoli15}, where emotions are induced by presenting diverse emotional scenes, while our study is specific to emotive faces. Prior neural studies~\cite{abadi2015decaf,Muhl14} which perform emotion recognition with lab-grade sensors achieve around 60\% VR. Superior performance of eye-based features can be attributed to the fact that the saccade and saliency-based statistics can capture discrepancies in gazing patterns for positive and negative facial emotions~\cite{Subramanian2014,Tavakoli15}; in contrast, neural features have only been moderately effective at decoding emotions conveyed by audio-visual music and movie stimuli~\cite{Koelstra2012,abadi2015decaf,Muhl14}, which can presumably elicit emotions more effectively than plain imagery.
 
\subsubsection{Gender Recognition}\label{GR_res}
GR involves labeling the test EEG or eye-gaze sample as arising from a \emph{male}/\emph{female} user. Table~\ref{tab:GenResults1} presents GR achieved with baseline classifiers, while Table~\ref{tab:GenResult2} shows AUC scores obtained with the EEG-based CNN, and the Adaboost ensemble fed with eye-gaze features. Both tables present GR results when user data corresponding to \textbf{All} emotional faces, and data for each of the six Ekman emotions (\textbf{Emotion-wise}) are employed for model training. Also, the best EF and LF results along with the relevant classifier are specified for the \textbf{All} condition. Cumulatively, Tables~\ref{tab:GenResults1} and~\ref{tab:GenResult2} clearly convey that the CNN and Adaboost frameworks considerably outperform baseline classifiers.

Focusing on Table~\ref{tab:GenResults1}, one can clearly note that the AUC scores with HI emotions are typically higher (cf. columns 1,2); also, AUC metrics achieved in the mouth-mask condition are typically higher than with eye-mask (columns 3,4). These findings from implicit user behavior mirror explicit recognition results in Fig.~\ref{RRnfix}. Specifically, significant differences can be noted with EEG-based results for the above conditions, while differences with eye-gaze are inconspicuous. These results cumulatively convey that (a) analyses of explicit and implicit user behavior affirm similar trends relating to visual FER; (b) gender differences are better encoded while perceiving high-intensity emotions, and how eye cues are interpreted for emotion inference, and (c) EEG signals better encode gender differences in visual emotion processing than gaze patterns.

Examining fusion results, LF is generally superior to EF when all emotions are considered, while EF achieves superior performance when emotion-specific user data are utilized for GR. We attempted GR from emotion-specific data to follow up on findings in Sec.~\ref{DA}, conveying females to be more sensitive to negative emotions. In Table~\ref{tab:GenResults1}, eye-gaze based GR performance improves significantly when emotion-specific data are considered (peak AUC of 0.68 for fear under eye-mask, vs peak AUC of 0.53 under mouth-mask for all emotions), while EEG results remain stable (peak AUC of 0.71 for HI anger vs peak AUC of 0.71 for all HI emotions). Optimal GR results (in bold) are mostly achieved for the A, D, F and Sa emotions implying that gender differences best manifest for negative valence.

Table~\ref{tab:GenResult2} largely replicates the trends noted from Table~\ref{tab:GenResults1}, with higher AUC scores achieved with the CNN and Adaboost classifiers. The one notable difference is that while the optimal NB baseline performs poorly with gaze features but much better with EEG in Table~\ref{tab:GenResults1}, the Adaboost framework outperforms the EEG-based CNN in the eye and mouth-mask conditions; EEG nevertheless encodes gender differences better while processing HI and LI emotions. These results imply that (a) eye movements in pursuit of FER, especially under partial face occlusion, are distinctive of gender, and (b) the discriminative Adaboost framework is able to better learn distinctive eye movement patterns as compared to the generative NB classifier. Late fusion results denoting the combination of decisions made by the CNN and Adaboost reveal that when one modality is considerably more potent than the other (EEG $>>$ Eye for HI and LI, while Eye $>$ EEG for mask), LF does not necessarily outperform constituent modalities.  

\begin{table*}[!t]
\fontsize{3}{3}\selectfont
\renewcommand{\tabcolsep}{5pt}
\centering
\caption{\textbf{GR} results with \textbf{baseline} classifiers over different modalities and their fusion.}\label{tab:GenResults1}
\resizebox{0.95\textwidth}{!}{
\begin{tabular}{l|l|c|cccc}
\hline
\multicolumn{3}{c|}{\textbf{AUC}} & \textbf{HI} & \textbf{LI} & \textbf{Eyemask} & \textbf{Mouthmask}\\
\hline
\multicolumn{2}{c|}{\multirow{4}{*}{\textbf{All}}} &
EEG~(NB) & \textbf{0.714 $\scriptscriptstyle{\pm}$ 0.002} & \textbf{0.600 $\scriptscriptstyle{\pm}$ 0.005} & \textbf{0.690 $\scriptscriptstyle{\pm}$ 0.03} & \textbf{0.654 $\scriptscriptstyle{\pm}$ 0.04} \\
\multicolumn{2}{c|}{} & EYE~(NB) & 0.493 $\scriptscriptstyle{\pm}$ 0.013 & 0.481 $\scriptscriptstyle{\pm}$ 0.017 & 0.471 $\scriptscriptstyle{\pm}$ 0.06 & 0.525 $\scriptscriptstyle{\pm}$ 0.05 \\
\multicolumn{2}{c|}{} & Early fusion~(RSVM) & 0.522 $\scriptscriptstyle{\pm}$ 0.035 & 0.524 $\scriptscriptstyle{\pm}$ 0.022 & 0.520 $\scriptscriptstyle{\pm}$ 0.07 & 0.520 $\scriptscriptstyle{\pm}$ 0.06 \\
\multicolumn{2}{c|}{} & Late fusion~(LSVM) & 0.549 $\scriptscriptstyle{\pm}$  0.022 & 0.523 $\scriptscriptstyle{\pm}$ 0.035 & 0.540 $\scriptscriptstyle{\pm}$ 0.05 & 0.610 $\scriptscriptstyle{\pm}$ 0.07 \\
\hline
\multirow{24}{*}{\rot{\textbf{Emotion wise}}} & \multirow{6}{*}{\rot{\textbf{EEG (NB)}}} & A &
\textbf{0.708 $\scriptscriptstyle{\pm}$ 0.064} & 0.580 $\pm $0.074 & 0.610 $\scriptscriptstyle{\pm}$ 0.16 & \textbf{0.672 $\scriptscriptstyle{\pm}$ 0.07} \\
 &  & D & 0.673 $\scriptscriptstyle{\pm}$ 0.055 & \textbf{0.696 $\scriptscriptstyle{\pm}$ 0.062} & 0.592 $\scriptscriptstyle{\pm}$ 0.06 & 0.650 $\scriptscriptstyle{\pm}$ 0.14 \\
 &  & F & 0.643 $\scriptscriptstyle{\pm}$ 0.059 & 0.596 $\scriptscriptstyle{\pm}$ 0.089 & 0.605 $\scriptscriptstyle{\pm}$ 0.14 & 0.564 $\scriptscriptstyle{\pm}$ 0.12 \\
 &  & H & 0.696 $\scriptscriptstyle{\pm}$ 0.047 & 0.668 $\scriptscriptstyle{\pm}$ 0.046 & 0.586 $\scriptscriptstyle{\pm}$ 0.07 & 0.624 $\scriptscriptstyle{\pm}$ 0.08 \\
 &  & Sa & 0.674 $\scriptscriptstyle{\pm}$ 0.048 & 0.634 $\scriptscriptstyle{\pm}$ 0.064 & \textbf{0.652 $\scriptscriptstyle{\pm}$ 0.08} & 0.590 $\scriptscriptstyle{\pm}$ 0.08 \\
 &  & Su & 0.692 $\scriptscriptstyle{\pm}$ 0.048 & 0.636 $\scriptscriptstyle{\pm}$ 0.071 & 0.633 $\scriptscriptstyle{\pm}$ 0.08 & 0.650 $\scriptscriptstyle{\pm}$ 0.09 \\
\cline{2-7} 
 & \multirow{6}{*}{\rot{\textbf{EYE (NB)}}} & 
A & \textbf{0.601 $\scriptscriptstyle{\pm}$ 0.021} & 0.565 $\scriptscriptstyle{\pm}$ 0.031 & 0.470 $\scriptscriptstyle{\pm}$ 0.20 & \textbf{0.590 $\scriptscriptstyle{\pm}$ 0.16} \\
 &  & D & 0.577 $\scriptscriptstyle{\pm}$ 0.011 & \textbf{0.632 $\scriptscriptstyle{\pm}$ 0.009}  & 0.480 $\scriptscriptstyle{\pm}$ 0.14 & 0.521 $\scriptscriptstyle{\pm}$ 0.27 \\
 &  & F & 0.595 $\scriptscriptstyle{\pm}$ 0.015 & 0.535 $\scriptscriptstyle{\pm}$ 0.029  & \textbf{0.680 $\scriptscriptstyle{\pm}$ 0.25} & 0.580 $\scriptscriptstyle{\pm}$ 0.17 \\
 &  & H & 0.560 $\scriptscriptstyle{\pm}$ 0.021 & 0.538 $\scriptscriptstyle{\pm}$ 0.017 & 0.555 $\scriptscriptstyle{\pm}$ 0.13 & 0.540 $\scriptscriptstyle{\pm}$ 0.12 \\
 &  & Sa & 0.539 $\scriptscriptstyle{\pm}$ 0.015 & 0.605 $\scriptscriptstyle{\pm}$ 0.030 & 0.445 $\scriptscriptstyle{\pm}$ 0.15 & 0.455 $\scriptscriptstyle{\pm}$ 0.13 \\
 &  & Su &0.555 $\scriptscriptstyle{\pm}$ 0.008 & 0.555 $\scriptscriptstyle{\pm}$ 0.018 & 0.624 $\scriptscriptstyle{\pm}$ 0.13 & 0.494 $\scriptscriptstyle{\pm}$ 0.16 \\
\cline{2-7} 
 & \multirow{6}{*}{\rot{\textbf{EF (RSVM)}}} & 
A & 0.555 $\scriptscriptstyle{\pm}$ 0.021 & 0.581 $\scriptscriptstyle{\pm}$ 0.037 & 0.320 $\scriptscriptstyle{\pm}$ 0.25 & 0.390 $\scriptscriptstyle{\pm}$ 0.18 \\
 & & D & 0.535 $\scriptscriptstyle{\pm}$ 0.024 & \textbf{0.622 $\scriptscriptstyle{\pm}$ 0.025} & 0.521 $\scriptscriptstyle{\pm}$ 0.18 & 0.500 $\scriptscriptstyle{\pm}$ 0.24 \\
 & & F & \textbf{0.618 $\scriptscriptstyle{\pm}$ 0.011} & 0.619 $\scriptscriptstyle{\pm}$ 0.041 & \textbf{0.700 $\scriptscriptstyle{\pm}$ 0.22} & 0.522 $\scriptscriptstyle{\pm}$ 0.18 \\
 & & H & 0.575 $\scriptscriptstyle{\pm}$ 0.017 & 0.597 $\scriptscriptstyle{\pm}$ 0.009 & 0.575 $\scriptscriptstyle{\pm}$ 0.13 & \textbf{0.524 $\scriptscriptstyle{\pm}$ 0.14} \\
 & & Sa & 0.540 $\scriptscriptstyle{\pm}$ 0.021 & 0.598 $\scriptscriptstyle{\pm}$ 0.024 & 0.532 $\scriptscriptstyle{\pm}$ 0.16 & 0.502 $\scriptscriptstyle{\pm}$ 0.14 \\
 & & Su & 0.579 $\scriptscriptstyle{\pm}$ 0.013 & 0.574 $\scriptscriptstyle{\pm}$ 0.014 & 0.433 $\scriptscriptstyle{\pm}$ 0.13 & 0.515 $\scriptscriptstyle{\pm}$ 0.16 \\
\cline{2-7} 
 & \multirow{6}{*}{\rot{\textbf{LF (RSVM)}}} &
A & 0.543 $\scriptscriptstyle{\pm}$ 0.062 & 0.571 $\scriptscriptstyle{\pm}$ 0.081 & 0.570 $\scriptscriptstyle{\pm}$ 0.15 & 0.590 $\scriptscriptstyle{\pm}$ 0.09 \\
 & & D & 0.542 $\scriptscriptstyle{\pm}$ 0.044 & \textbf{0.597 $\scriptscriptstyle{\pm}$ 0.093} & 0.590 $\scriptscriptstyle{\pm}$ 0.10 & 0.570 $\scriptscriptstyle{\pm}$ 0.14 \\
 & & F & 0.519 $\scriptscriptstyle{\pm}$ 0.029 & 0.597 $\pm $ 0.170 & \textbf{0.645 $\scriptscriptstyle{\pm}$ 0.16} & \textbf{0.691 $\scriptscriptstyle{\pm}$ 0.12} \\
 & & H & 0.526 $\scriptscriptstyle{\pm}$ 0.031 & 0.508 $\pm $0.017 & 0.564 $\scriptscriptstyle{\pm}$ 0.09 & 0.552 $\scriptscriptstyle{\pm}$ 0.10 \\
 & & Sa & \textbf{0.573 $\scriptscriptstyle{\pm}$ 0.076} & 0.584 $\scriptscriptstyle{\pm}$ 0.102 & 0.513 $\scriptscriptstyle{\pm}$ 0.04 & 0.580 $\scriptscriptstyle{\pm}$ 0.11 \\
 & & Su & 0.562 $\scriptscriptstyle{\pm}$ 0.073 & 0.568 $\scriptscriptstyle{\pm}$ 0.125 & 0.581 $\scriptscriptstyle{\pm}$ 0.08 & 0.583 $\scriptscriptstyle{\pm}$ 0.08 \\
\hline
\end{tabular}}
\vspace{0.2cm}
\renewcommand{\tabcolsep}{5pt}
\centering
\caption{\textbf{GR} results with \textbf{CNN} and \textbf{Adaboost} classifiers over different modalities and their fusion.}\label{tab:GenResult2}
\resizebox{0.95\textwidth}{!}{
\begin{tabular}{l|l|c|cccc}
\hline
\multicolumn{3}{c|}{\textbf{AUC}} & \textbf{HI} & \textbf{LI} & \textbf{eye mask} & \textbf{mouth mask}\\
\hline
\multicolumn{2}{c|}{\multirow{4}{*}{\textbf{All}}} &
EEG~(CNN) & \textbf{0.931 $\scriptscriptstyle{\pm}$ 0.011} & \textbf{0.882 $\scriptscriptstyle{\pm}$ 0.020} & 0.892 $\scriptscriptstyle{\pm}$ 0.038 & 0.886 $\scriptscriptstyle{\pm}$ 0.023 \\
\multicolumn{2}{c|}{} & EYE~(Ada) & 0.521 $\scriptscriptstyle{\pm}$ 0.016 & 0.538 $\scriptscriptstyle{\pm}$ 0.023 & 0.969 $\scriptscriptstyle{\pm}$ 0.015 & \textbf{0.966 $\scriptscriptstyle{\pm}$ 0.014} \\
\multicolumn{2}{c|}{} & Late fusion & 0.920 $\scriptscriptstyle{\pm}$  0.015 & 0.850 $\scriptscriptstyle{\pm}$ 0.045 & \textbf{0.974 $\scriptscriptstyle{\pm}$ 0.017} & 0.946 $\scriptscriptstyle{\pm}$ 0.037 \\
\hline

\multirow{18}{*}{\rot{\textbf{Emotion wise}}} & \multirow{6}{*}{\rot{\textbf{EEG (CNN)}}} & A &
0.757 $\scriptscriptstyle{\pm}$ 0.113 & 0.719	$\scriptscriptstyle{\pm}$ 0.086 & 0.626 $\scriptscriptstyle{\pm}$ 0.165 & 0.676 $\scriptscriptstyle{\pm}$ 0.157 \\
 &  & D & 0.758 $\scriptscriptstyle{\pm}$ 0.047 & 0.644 $\scriptscriptstyle{\pm}$ 0.080 & 0.594 $\scriptscriptstyle{\pm}$ 0.161 & 0.673 $\scriptscriptstyle{\pm}$ 0.108 \\
 &  & F & 0.738 $\scriptscriptstyle{\pm}$ 0.083 & 0.618 $\scriptscriptstyle{\pm}$ 0.107 & 0.542 $\scriptscriptstyle{\pm}$ 0.128 & 0.585 $\scriptscriptstyle{\pm}$ 0.199  \\
 &  & H & \textbf{0.790 $\scriptscriptstyle{\pm}$ 0.049} & 0.701 $\scriptscriptstyle{\pm}$ 0.054 & \textbf{0.714 $\scriptscriptstyle{\pm}$ 0.082} & \textbf{0.751 $\scriptscriptstyle{\pm}$ 0.051} \\
 &  & Sa & 0.739 $\scriptscriptstyle{\pm}$ 0.089 &\textbf{0.737 $\scriptscriptstyle{\pm}$ 0.078} & 0.689 $\scriptscriptstyle{\pm}$ 0.097 & 0.731 $\scriptscriptstyle{\pm}$ 0.079 \\
 &  & Su & 0.720 $\scriptscriptstyle{\pm}$ 0.076 & 0.636 $\scriptscriptstyle{\pm}$ 0.042 & 0.649 $\scriptscriptstyle{\pm}$ 0.098 & 0.652 $\scriptscriptstyle{\pm}$ 0.067 \\
\cline{2-7} 

& \multirow{6}{*}{\rot{\textbf{EYE (Ada)}}} & 
A & \textbf{0.651 $\scriptscriptstyle{\pm}$ 0.064} & \textbf{0.651 $\scriptscriptstyle{\pm}$ 0.079} & 0.840 $\scriptscriptstyle{\pm}$ 0.160 & 0.957 $\scriptscriptstyle{\pm}$ 0.038 \\
 &  & D & 0.591 $\scriptscriptstyle{\pm}$ 0.038 & 0.600 $\scriptscriptstyle{\pm}$ 0.068  & 0.950 $\scriptscriptstyle{\pm}$ 0.045 & 0.910 $\scriptscriptstyle{\pm}$ 0.108 \\
 &  & F & 0.569 $\scriptscriptstyle{\pm}$ 0.078 & 0.528 $\scriptscriptstyle{\pm}$ 0.157  & 0.950 $\scriptscriptstyle{\pm}$ 0.067 & 0.881 $\scriptscriptstyle{\pm}$ 0.075 \\
 &  & H & 0.574 $\scriptscriptstyle{\pm}$ 0.090 & 0.529 $\scriptscriptstyle{\pm}$ 0.050 & 0.931 $\scriptscriptstyle{\pm}$ 0.049 & 0.929 $\scriptscriptstyle{\pm}$ 0.051 \\
 &  & Sa & 0.587 $\scriptscriptstyle{\pm}$ 0.087 & 0.576 $\scriptscriptstyle{\pm}$ 0.068 & \textbf{0.959 $\scriptscriptstyle{\pm}$ 0.051} & \textbf{0.958 $\scriptscriptstyle{\pm}$ 0.041} \\
 &  & Su &0.530 $\scriptscriptstyle{\pm}$ 0.093 & 0.546 $\scriptscriptstyle{\pm}$ 0.052 & 0.938 $\scriptscriptstyle{\pm}$ 0.067 & 0.921 $\scriptscriptstyle{\pm}$ 0.107 \\
\cline{2-7} 
& \multirow{6}{*}{\rot{\textbf{LF}}} & 
A & 0.720 $\scriptscriptstyle{\pm}$ 0.082 & 0.651 $\scriptscriptstyle{\pm}$ 0.113 & 0.835 $\scriptscriptstyle{\pm}$ 0.167 & 0.860 $\scriptscriptstyle{\pm}$ 0.132 \\
 & & D & 0.676 $\scriptscriptstyle{\pm}$ 0.083 & 0.557 $\scriptscriptstyle{\pm}$ 0.153 & \textbf{0.939 $\scriptscriptstyle{\pm}$ 0.070} & \textbf{0.967 $\scriptscriptstyle{\pm}$ 0.105} \\
 & & F & 0.675 $\scriptscriptstyle{\pm}$ 0.140 & 0.470 $\scriptscriptstyle{\pm}$ 0.164 & 0.887 $ \pm $ 0.139 & 0.869 $\scriptscriptstyle{\pm}$ 0.097 \\
 & & H & \textbf{0.770 $\scriptscriptstyle{\pm}$ 0.050} & \textbf{0.683 $\scriptscriptstyle{\pm}$ 0.088} & 0.897 $\scriptscriptstyle{\pm}$ 0.042 & 0.926 $\scriptscriptstyle{\pm}$ 0.060 \\
 & & Sa & 0.729 $\scriptscriptstyle{\pm}$ 0.062 & 0.555 $\scriptscriptstyle{\pm}$ 0.107 & \textbf{0.939 $\scriptscriptstyle{\pm}$ 0.062} & 0.843 $\scriptscriptstyle{\pm}$ 0.133 \\
 & & Su & 0.723 $\scriptscriptstyle{\pm}$ 0.090 & 0.604 $\scriptscriptstyle{\pm}$ 0.105 & 0.826 $\scriptscriptstyle{\pm}$ 0.142 & 0.943 $\scriptscriptstyle{\pm}$ 0.058 \\
\cline{2-7} 
\hline
\end{tabular}}
\vspace{-.4cm}
\label{GenDeep}
\end{table*}

\subsubsection{Spatio-temporal EEG analysis} 
We also examined spatio-temporal characteristics of the EEG signal to examine if (i)  gender differences in visual emotion processing are effectively captured by certain electrodes, and are attributable to specific (functional) brain lobes and (ii) any time window(s) were critical for GR over 4s of visual processing. \\

\noindent \textbf{Spatial:} We evaluated the ability of each EEG channel (cf. Fig~\ref{Emotiv}) to capture gender-discriminative information by feeding the deep network (Fig.~\ref{DLarch}) with single-channel EEG input.  Consistent with prior findings~\cite{maneesh2017icmi,maneesh2017acii,Mclure04}, we noted optimal GR with the \emph{frontal} and \emph{occipital} brain lobes. Single channel GR followed a trend similar to Table~\ref{GenDeep}, with optimal GR achieved for HI emotions, and worst GR performance noted with mouth-mask data. A symmetricity was noted among the optimal EEG channels, namely, AF3 and AF4, and F3 and F4. These results mirror observations relating to the existence of \emph{brain hemispheres}\cite{vingerhoets2003cerebral} for ER. \\

\begin{figure*}[th!]
\centering
\includegraphics[width=1.0\linewidth]{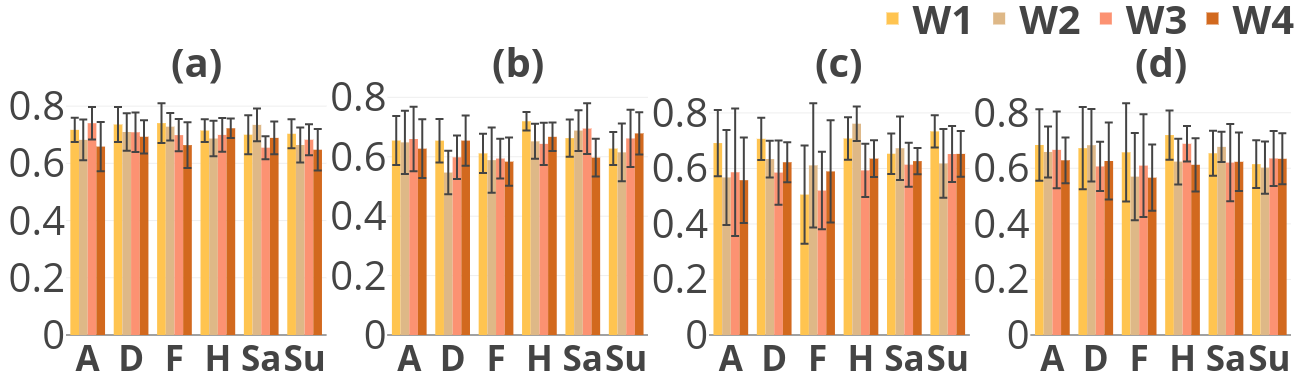}
\caption{(a--d) \textbf{Temporal EEG analyses}: GR for {HI}, {LI}, {Eye} and {Mouth mask} conditions over temporal windows (W1--W4).} 
\label{winGenResults}
\vspace{-.4cm}
\end{figure*}

\noindent \textbf{Temporal:} As isolation of gender-specific ERPs is possible from 1s time-windows (Sec.~\ref{ERP}), we considered four non-overlapping 1s windows W1--W4 spanning 4s of stimulus viewing. Fig.~\ref{winGenResults} presents GR AUCs achieved over W1--W4 from emotion-specific EEG data. Plots confirm that reliable, above-chance GR is achieved over each of W1--W4 across the different viewing conditions. Highest GR performance is noted for HI morphs and lowest GR for {eye mask} faces, consistent with Table~\ref{GenDeep}. While temporal analyses revealed no significant GR differences across windows, optimal GR was generally achieved for W1 and W2 over all morph conditions suggesting a \emph{primacy} effect. Masked conditions result in higher GR performance variance, perhaps due to the initial difficulty in FER owing to facial occlusions.

\begin{table}[]
\centering
\caption{\textbf{Spatial performance} evaluation. Parentheses denote channel for which best AUC was achieved.} \vspace{-.1cm}
\fontsize{6.5}{6.5}\selectfont
\label{temporal_table}
\begin{tabular}{lllll}
\hline
\hline


 & HI               & LI                & Eye-mask          & Mouth-mask        \\ \hline
1                 & \textbf{0.8266 (AF4)} & 0.8008 (F3) & 0.7160 (F4) & 0.7135 (F3) \\ 
2                 & 0.8200 (F3) & 0.7772 (O2) & 0.7143 (F3) & 0.7124 (F4) \\
3                 & 0.8131 (O2)  & 0.7745 (AF4) & 0.7055 (AF4) & 0.7107 (AF3)\\
\hline
\hline
\end{tabular}
\vspace{-.4cm}
\end{table}


\section{Discussion \& Conclusion}\label{Con}
A critical requirement of today's ubiquitous computing devices is to sense user intention, emotion and cognition from multimodal cues, and devise effective interactions to optimize users' individual and social behaviors. Affective computing (AC), whose objective is to empower devices to interact naturally and empathetically with users, plays a crucial role here. Apart from inferring user emotions, it would also be beneficial for AC systems to predict \emph{soft biometrics} such as user \emph{age} and \emph{gender} for effective interaction~\cite{Rukavina16}. There is also a strong need to develop sensing mechanisms that respect user privacy concerns~\cite{Reynolds04}, and eye-movements and EEG signals represent \emph{privacy preserving implicit behaviors} which enable inference of user traits even as the user identity remains hidden. 

The primary objective of this study is to explore the utility of eye movements and EEG for \emph{user gender} prediction. We designed a study where these implicit user behaviors were recorded as 28 users (14 male) recognized facial emotions; since eye movements are known to be characteristic of the perceived facial emotion~\cite{schurgin,subramanian2011can}, our study design additionally enabled recognition of the \emph{facial} (stimulus) \emph{valence}. Crucially, our study employs lightweight and inexpensive sensors for inference, as against bulky and intrusive lab sensors which are typically used in user-centered analyses but significantly constrain user behavior. Also, to examine user efficacy for FER under occlusions, we presented both unoccluded (high and low-intensity), and occluded (with the eye or mouth region masked) emotive faces to users.

The fact that gender differences exist in visual emotional processing is demonstrated at multiple levels via our experiments. In terms of \emph{explicit} user responses, female users are found to quickly and accurately perform FER on unoccluded faces, especially for negative valence emotions (Section~\ref{DA}). Likewise, females also achieve higher recognition of negative emotions with mouth-masked faces. Subsequent examination of \emph{implicit} cues revealed interesting correlations; examination of eye fixation distributions showed greater female fixation around the eyes than males irrespective of stimulus type, and fixations on eyes were significantly longer in the mouth-mask condition. This observation is consistent with prior studies~\cite{wells2016identification}, and the proposition that females primarily look at the eyes for emotional cues.     

Analysis of EEG ERPs also conveyed interesting patterns. While processing unoccluded faces, stronger N100 and P300 peaks are noted in female ERPs for negative emotions, as well as a stronger N400 peak for strongly negative faces. This suggests a differential cognitive processing of negative vs positive emotions by females, which results in their enhanced sensitivity towards negative emotions. Likewise, lower male N400 latencies are generally noted at the O2 electrode for positive emotions. Stronger N100 and P300 female ERPs are also noted for negative mouth-masked emotions, suggesting that negative emotions are processed rapidly by women, and do not  necessarily entail selective attention to emotional cues. 	 	  
  
That implicit and differential behaviors can be isolated using data acquired via low-cost sensors affirms that our experimental design can effectively capture emotion and gender-specific information. Greater female sensitivity to negative emotions can be attributed to several factors like social structure, environment and evolution, way of living and social stereotypes\cite{deng2016gender}. While we did not seek to expressly elicit emotions through facial imagery, facial expressions are known to induce emotions in the viewer~\cite{Siedlecka19}, and examination of users' emotional behavior enables prediction of both the {user gender} and {stimulus valence}.

Valence recognition experiments employing eye-gaze features (processed by an Adaboost ensemble) and EEG features (input to a 3-layer CNN performing 1D convolutions) revealed the following. Eye-gaze patterns on HI emotional faces were highly characteristic of facial valence for both male and female users, resulting in AUC scores $\geq$ 0.98. Contrastingly, EEG produced largely near-chance performance, implying that the EEG features are sub-optimal for ER. Gulf in the efficacy of eye-gaze and EEG features meant that late fusion of the classifier outputs was hardly beneficial. That human eye movements are highly characteristic of stimulus valence is unsurprising, with prior studies~\cite{Tavakoli15} achieving better-than-chance accuracy with gaze features compiled for diverse scenes; on the contrary, our study is specific to emotional faces.

Gender recognition results are summarized as follows. Table~\ref{tab:GenResults1} presenting GR results achieved with the baseline NB, LSVM and RSVM classifiers affirms that consistent with the recognition rate statistics and ERP analyses, EEG-based gender differences best manifest for HI emotional faces, and are least observable for LI emotional faces. Emotion-specific analyses affirm that peak GR AUC scores are mostly achieved with negative valence data. Also, the NB classifier achieves much superior results with EEG as compared to eye-gaze features. On the other hand, Table~\ref{tab:GenResult2} shows that much superior GR results are achievable with the CNN and Adaboost classifiers respectively fed with the EEG and eye-gaze data. Interestingly, very high AUC scores are achieved with eye-gaze features compiled for the mask conditions (this trend is unobservable for the HI and LI conditions), implying that eye movements in pursuit of FER under partial face occlusion are highly gender-specific. The substantial disparity in EEG and eye-based results across the different conditions results in the fusion of the two modalities is hardly beneficial. 	  

While the presented study involves only a small pool of (N=28) users, the observed GR and ER results are nevertheless highly promising and demonstrate the utility of implicit behaviors for privacy preserving user profiling. The ever-increasing and commonplace availability of sensors employed in this work also opens up the possibility of conducting large-scale (crowdsourced) user-centric studies. Our larger endeavor is to predict \emph{soft biometrics} (age, gender, emotional and cognitive state) of users via implicit and multimodal behavioral cues to empower gaming, advertising, augmented and virtual reality applications for behavioral change, and mental health monitoring systems for disorders like Alexithymia. Future work will focus on the modeling of shared behaviors (joint embedding of EEG and eye-gaze features) for efficient user-trait prediction employing techniques such as multi-task learning, and prototyping real-life profiling applications. Investigating the optimality of pre-designed features (\eg, power-spectral density for EEG) or learned feature descriptors represents another interesting line of work.

\ifCLASSOPTIONcaptionsoff
  \newpage
\fi



%

\bibliographystyle{IEEEtran}
\bibliography{short,bibfile}

%
\vspace{-.5in}
\begin{IEEEbiography}[{\includegraphics[width=1in,height=1.25in,clip,keepaspectratio]{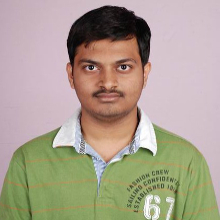}}]{Maneesh Bilalpur}
is a Ph.D. student at the University of Pittsburgh. He holds a Bachelor's degree in Electronics and Communication Engineering from the Vellore Institute of Technology, and a Master of Science by Research from the Center for Visual Information Technology at International Institute of Information Technology Hyderabad, India. His research interests include Machine Learning, Affective Computing, Multimedia Analysis and Computer Vision.
\end{IEEEbiography}
\vspace{-.5in}
\begin{IEEEbiography}[{\includegraphics[width=1in,height=1.25in,clip,keepaspectratio]{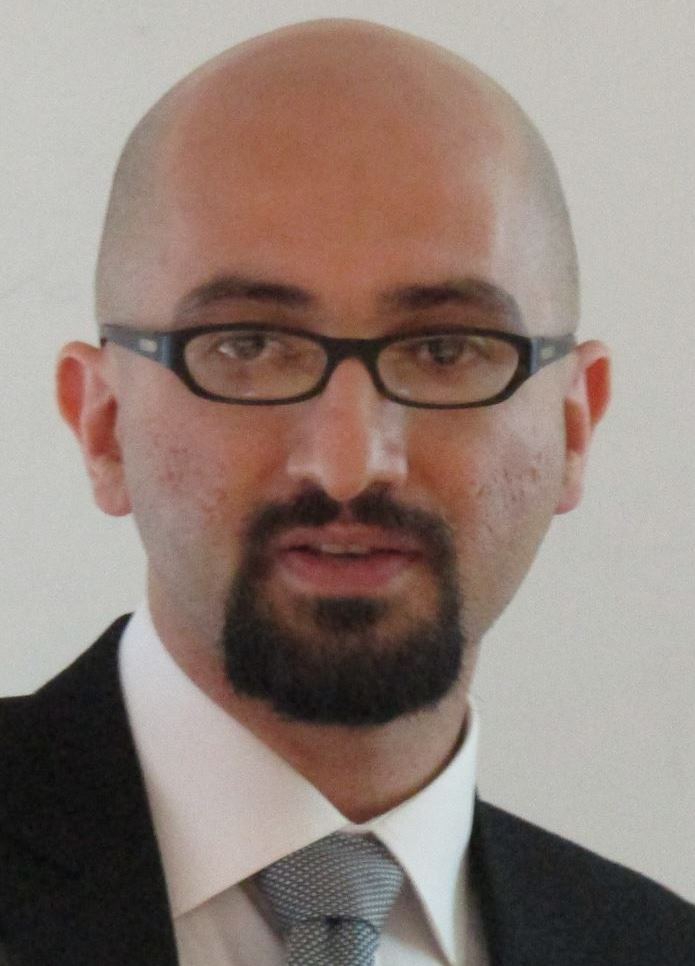}}]{Seyed Mostafa Kia}
received his PhD degree in Computer Science from the University of Trento, Italy. His research focused on the interpretability of MEG decoding models. He is now a post doctoral researcher at the Radboud university medical center in Nijmegen, Netherlands. He seeks scalable machine learning solutions to solve intractable problems in clinical neuroimaging data analysis. 
\end{IEEEbiography}
\vspace{-.5in}
\begin{IEEEbiography}[{\includegraphics[width=1in,height=1.25in,clip,keepaspectratio]{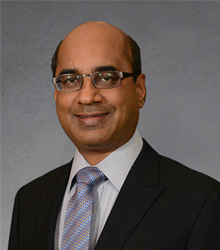}}]{Mohan Kankanhalli} is a Professor and Dean of the School of Computing at the National University of Singapore (NUS). Mohan obtained his BTech from IIT Kharagpur and MS and PhD from Rensselaer Polytechnic. His current research interests are in Multimedia Systems (content processing, retrieval) and Multimedia Security (surveillance and privacy). Mohan is actively involved in organizing major conferences, and serves on the editorial board of several journals.
\end{IEEEbiography}
\vspace{-.5in}
\begin{IEEEbiography}[{\includegraphics[width=1in,height=1.25in,clip,keepaspectratio]{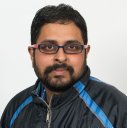}}]{Ramanathan Subramanian}
received his Ph.D. in Electrical and Computer Engg. from NUS in 2008. He is an Associate Professor in Computer Science and Engg. at IIT Ropar. His past affiliations include IHPC (Singapore), U Glasgow (Singapore), IIIT Hyderabad (India) and UIUC-ADSC (Singapore). His research focuses on Human-centered computing, and especially on extracting and modeling non-verbal behavioral cues for interactive analytics. He is an IEEE Senior Member and a member of the ACM and AAAC.
\end{IEEEbiography}

\end{document}